\documentclass[twocolumn,english,aps,manuscript,prl,reprint,longbibliography]{revtex4-1}
\usepackage[T1]{fontenc}
\usepackage[latin9]{inputenc}
\usepackage{amsmath}
\usepackage{amssymb}
\usepackage{graphicx}
\usepackage{natbib}
\usepackage{babel}
\usepackage{soul}

\makeatletter
\usepackage[usenames,dvipsnames]{color}
 \@ifundefined{textcolor}{}
 {%
   \definecolor{BLUE}{rgb}{0,0,1}
   \definecolor{RED}{rgb}{ 0.6328125,0,0}
 }

\newcommand{\ket}[1]{\left|#1\right>}

\newcommand{\eqtext}[1]{\text{\textnormal{#1}}}
\DeclareMathOperator{\Tr}{Tr}
\makeatother

\begin{document}

\title{Ergodic dynamics and thermalization in an isolated quantum system}

\author{C. Neill$^{1}$}
\thanks{These authors contributed equally to this work.}
\author{P. Roushan$^{2}$}
\thanks{These authors contributed equally to this work.}
\author{M. Fang$^{1}$}
\thanks{These authors contributed equally to this work.}
\author{Y. Chen$^{2}$}
\thanks{These authors contributed equally to this work.}
\author{M. Kolodrubetz$^{3}$}

\author{Z. Chen$^{1}$}
\author{A. Megrant$^{2}$}
\author{R. Barends$^{2}$}
\author{B. Campbell$^{1}$}
\author{B. Chiaro$^{1}$}
\author{A. Dunsworth$^{1}$}
\author{E. Jeffrey$^{2}$}
\author{J. Kelly$^{2}$}
\author{J. Mutus$^{2}$}
\author{P. J. J. O'Malley$^{1}$}
\author{C. Quintana$^{1}$}
\author{D. Sank$^{2}$}
\author{A. Vainsencher$^{1}$}
\author{J. Wenner$^{1}$}
\author{T. C. White$^{2}$}
\author{A. Polkovnikov$^{3}$}
\author{J. M. Martinis$^{1,2}$}

\email{martinis@physics.ucsb.edu}

\affiliation{$^{1}$Department of Physics, University of California, Santa Barbara,
CA 93106-9530, USA}

\affiliation{$^{2}$Google Inc., Santa Barbara,
CA 93117, USA}

\affiliation{$^{3}$Department of Physics, Boston University, Boston,
MA 02215, USA}

\maketitle

\textbf{Statistical mechanics is founded on the assumption that all accessible configurations of a system are equally likely. 
This requires dynamics that explore all states over time, known as ergodic dynamics.
In isolated quantum systems, however, the occurrence of ergodic behavior has remained an outstanding question \cite{deutsch1991quantum, srednicki1994chaos,  rigol2008thermalization, polkovnikov2011}.
Here, we demonstrate ergodic dynamics in a small quantum system consisting of only three superconducting qubits.
The qubits undergo a sequence of rotations and interactions and we measure the evolution of the density matrix.
Maps of the entanglement entropy show that the full system can act like a reservoir for individual qubits, increasing their entropy through entanglement.
Surprisingly, these maps bear a strong resemblance to the phase space dynamics in the classical limit; classically chaotic motion coincides with higher entanglement entropy.
We further show that in regions of high entropy the full multi-qubit system undergoes ergodic dynamics.
Our work illustrates how controllable quantum systems can investigate fundamental questions in non-equilibrium thermodynamics.}

%\psection{Introduction.}
Imagine air molecules in a room.  
They move around with all possible velocities in all directions.
Attaining the exact knowledge of these trajectories is a daunting and an unrealistic task.  
Statistical mechanics, however, claims that exact knowledge of individual trajectories is not required and systems can be accurately described using only a few parameters.
What is the essential property of these systems that allows for such a simple description?

Ergodic dynamics provide an explanation for this simplicity.
If the dynamics are ergodic, then the system will uniformly explore all microscopic states over time, constrained only by conservation laws.
Ergodicity ensures that
\begin{equation}\label{Ergodic}
\left<O\right>_{\eqtext{time}} = \left<O\right>_{\eqtext{states}}
\end{equation}
where $O$ is any macroscopic observable and brackets denote averaging.
In thermal equilibrium, observables are stationary and therefore at all times $O\left(t\right) = \left<O\right>_{\eqtext{time}}$.
These two equations imply that, at all times, observables are given by an average over states and this forms the foundation for all thermodynamic calculations.

In classical systems, it is chaotic motion which drives the system to ergodically explore the state space \cite{ott2002chaos}.
Quantum systems, however, are governed by Schrodinger's equation which is linear and consequently forbids chaotic motion \cite{gutzwiller1990chaos}.
This poses fundamental questions regarding the applicability of statistical mechanics in isolated quantum systems \cite{deutsch1991quantum, srednicki1994chaos,  rigol2008thermalization, polkovnikov2011}.
Do quantum systems exhibit ergodic behavior in the sense of Eq.\,\ref{Ergodic}?
Do quantum systems act as their own bath in order to approach thermal equilibrium?
Extensive experimental efforts have been made to address these fundamental questions \cite{kinoshita2006quantum,klaers2010thermalization,cheneau2012light,trotzky2012probing,langen2013local,richerme2014non,schreiber2015observation}.

%\psection{Theory.}
Here we investigate ergodic dynamics by considering a simple quantum model whose classical limit is chaotic \cite{wang2004entanglement, ghose2008chaos, lombardi2011entanglement, chaudhury2009quantum, haake1987classical}.
This model describes a collection of spin-1/2 particles whose collective motion is equivalent to that of a single larger spin with total angular momentum $j$ governed by the Hamiltonian
\begin{equation}\label{Hamiltonian}
\mathcal{H}\left( t \right) = \frac{\pi}{2 \tau} J_y + \frac{\kappa}{2j}  J_z^2 \sum_{n=1}^N \delta \left( t - n\tau \right)
\end{equation}
where $J_y$ and $J_z$ are angular momentum operators.
The sum over delta functions implies $N$ applications of $J_z^2$ each at integer time steps.
The angular momentum operators can be expressed in terms of the constituent spin-1/2 Pauli operators, e.g. $J_z = \frac {\hbar} {2} \sum\nolimits_{i} {z_i}$.
Setting $\tau = 1$, the first term in $\mathcal{H}$ causes each spin to rotate around the $y$-axis by an angle $\pi/2$.
The second term couples every spin to every other spin with strength $\kappa/2j$.
This can be seen by expanding $J_z^2$ in terms of $z$ operators, where terms like $z_1 z_2$ and all other combinations appear.

The classical dynamics, being simple to visualize and interpret, can provide valuable intuition for studying the quantum limit.
The classical limit of this model occurs when $j$ is very large and quantization effects become negligible.
In this limit, the system behaves like a classical spinning top with dynamics which are known to be chaotic \cite{wang2004entanglement, ghose2008chaos, lombardi2011entanglement, chaudhury2009quantum}.
The parameter $\kappa$ sets the chaoticity and takes the dynamics from regular to chaotic as $\kappa$ increases; at intermediate values, the system exhibits a rich mixture of both regular and chaotic motion.

\begin{figure}[t!]
\includegraphics{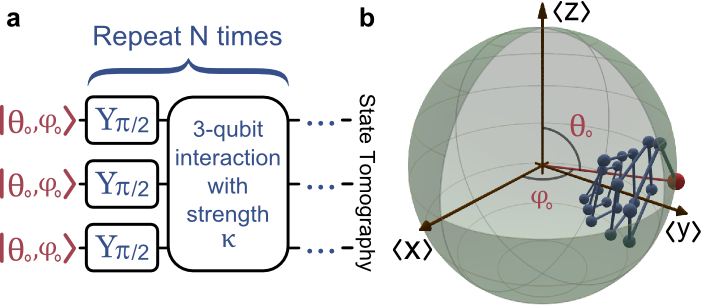} 
\caption {
\textbf{Pulse sequence and the resulting quantum dynamics. a,}  Pulse sequence showing first the initial state of the three qubits (Eq.\,\ref{CoherentState}) followed by the unitary operations for a single time step (Eq.\,\ref{UnitaryEvolution}).  These operations are repeated $N$ times before measurement.  Single qubit rotations are generated using shaped-microwave pulses in 20\,ns; the three-qubit interaction is generated using a tunable coupling circuit controlled using square pulses of length 5\,ns for $\kappa = 0.5$ and 25\,ns for $\kappa = 2.5$. \textbf{b,} The state of a single qubit is measured using state tomography and shown in a Bloch sphere. The initial state is shown in red with subsequent states shown in blue for $N = 1$ to $20$.
}
\label{fig:PulseSequence} 
\end{figure}

\begin{figure*}
\begin{centering}
\includegraphics[width=184mm]{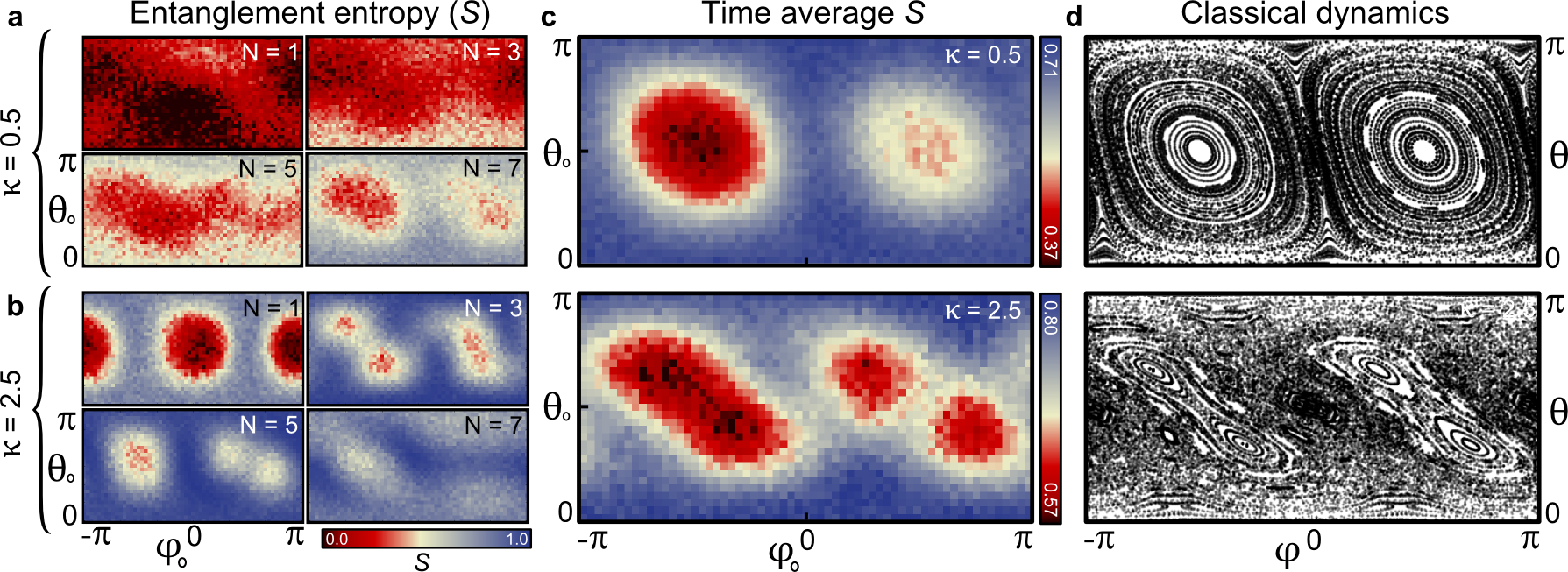}
\par\end{centering}
\caption{
\textbf{Entanglement entropy and classical chaos. a,b,} The entanglement entropy (color) of a single qubit (see Eq.\,\ref{EntanglementEntropy}) averaged over qubits and mapped over a 31\,x\,61 grid of initial state, for various time steps $N$ and two values of interaction strength $\kappa$. The entanglement entropy of a single qubit can range from 0 to 1.  \textbf{c,} The entanglement entropy averaged over 20 steps for $\kappa = 0.5$ and over 10 steps for $\kappa = 2.5$; for both experiments the maximum pulse sequence is $\approx500$\,ns.  The left/right asymmetry is the result of experimental imperfections and is not present in numerical simulations (see supplement). \textbf{d,} A stroboscopic map of the classical dynamics is computed numerically and shown for comparison.  The map is generated by randomly choosing initial states, propagating each state forward using the classical equations of motion, and plotting the orientation of the state after each step as a point.  We observe a clear connection between regions of chaotic behavior (classical) and high entanglement entropy (quantum).
}
\label{fig:PhaseSpace}
\end{figure*}

%\psection{Figure 1.}
Experimentally realizing this model requires a high degree of control over both local terms and interactions in a multi-qubit Hamiltonian.
This led to the design of a three-qubit ring of planar transmons with tunable inter-qubit coupling (see supplement) \cite{barends2013coherent, chen2014qubit, geller2014tunable}.
The rotations around the $y$-axis ($J_y$) are performed using shaped microwave pulses that are resonant with the qubit transition.
The simultaneous and symmetric three-qubit interaction ($J_z^2$) is turned on and off using a tunable coupling circuit controlled by three separate square pulses.
The qubit-qubit interaction energy $g$ and the duration of the interaction pulses $T$ set $\kappa$ through the relation $\kappa = 3gT/\hbar$.
We measure the strength of the interaction energy $\kappa$ by determining the time it takes for an excitation to swap between the qubits (see supplement).

The periodic nature of $\mathcal{H}$ allows us to write down the unitary evolution over one cycle as
\begin{equation}\label{UnitaryEvolution}
U = e^{-i \frac{\kappa}{2j\hbar}  J_z^2} e^{-i \frac{\pi}{2 \hbar} J_y}
\end{equation}
shown schematically in Fig.\,\ref{fig:PulseSequence}a.
We begin by initializing each qubit in the state
\begin{equation}\label{CoherentState}
\ket{\theta_0, \varphi_0} = \cos\left(\theta_0\right) \ket{+z} + e^{-i\varphi_0} \sin\left(\theta_0\right) \ket{-z}
\end{equation}
where $\theta_0$ and $\varphi_0$ are angles describing the orientation of the single qubit states.
This state is known as a spin coherent state and is the most classical spin state in the sense of minimum uncertainty and zero entanglement.
We then rotate each qubit around the $y$-axis by $\pi/2$, followed by a simultaneous multi-qubit interaction.
We repeat these two operations $N$ times and then tomographically reconstruct the resulting density matrix \cite{neeley2010generation}.
For details regarding the pulse sequence see supplementary information.

We visualize the evolution of the system by depicting the single-qubit state as a vector inside of a Bloch sphere, shown in Fig.\,\ref{fig:PulseSequence}b.
Each Bloch vector is constructed by measuring the expectation values of the $x$, $y$, and $z$ Pauli operators after evolving the system according to Eq.\,\ref{UnitaryEvolution}.
As the dynamics are symmetric under qubit exchange, the qubits undergo nominally identical evolution and we plot the average behavior (see supplement).
The chosen initial state is shown in red with the Bloch vector after subsequent steps shown in blue.
After each step, there are two qualitative changes:  a rotation and a change in the length.
The orientation is analogous to the orientation of the classical spin.
The change in length, however, describes entanglement amongst the qubits.

%\psection{Figure 2.}
Entanglement can be characterized using the entanglement entropy $S$,
\begin{equation}\label{EntanglementEntropy}
S = - \Tr{\rho_{\eqtext{sq}} \log_2\left(\rho_{\eqtext{sq}}\right)}
\end{equation}
where $\rho_{\eqtext{sq}}$ is the density matrix of a single qubit.
Writing the trace as a sum reproduces the familiar definition of entropy $-\sum{p_i \log\left(p_i\right)}$, where $p_i$ is the probability of being in the $i$th microstate.
If the qubit is in a pure state, then the single-qubit state is completely known and the entropy is zero.
However, if the qubits are entangled with one another, then $\rho_{\eqtext{sq}}$ is a statistical mixture of states and the entropy is non-zero.

In  Fig.\,\ref{fig:PhaseSpace}a we show the entanglement entropy between a single qubit and the rest of the qubits at several instances in time.
In each panel we prepare various initial states $\ket{\theta_0, \varphi_0}$, evolve the system for $N$ steps and plot the entanglement entropy; different panels correspond to different $N$.
Initial states prepared close to the $y$-axis have low entropy (red) which remains low as the system evolves.
States prepared farther away from the $y$-axis gain higher entropy (blue) given sufficient time.
We perform the same set of experiments for stronger interaction, $\kappa = 2.5$, shown in Fig.\,\ref{fig:PhaseSpace}b.
At stronger interactions, the entropy increases rapidly and regions of low entropy are no longer isolated to near the $y$-axis.

In  Fig.\,\ref{fig:PhaseSpace}\,a,b we see that the entropy fluctuates over time.
In small quantum systems, there are fluctuations or revivals that vanish when the system size is taken to infinity (known as the thermodynamic limit).
For finite systems, averaging the entropy over time is commonly used to estimate the equilibrium value approached by larger systems.
In Fig.\,\ref{fig:PhaseSpace}c we show the entanglement entropy averaged over time ($N$) for both values of interaction strength $\kappa$.
The corresponding classical dynamics are shown in Fig.\,\ref{fig:PhaseSpace}d.

We find a striking resemblance between entanglement in the quantum system and chaotic dynamics in the classical limit.
The regions of classical phase space where the dynamics are chaotic correspond to high entropy (blue) in the quantum system; regions that are classically regular correspond to low entropy (red), including bifurcation at large $\kappa$.
The results shown in Fig.\,\ref{fig:PhaseSpace}b have been predicted by recent theoretical works \cite{khripkov2013coherence, madhok2013signatures}.
However, these studies focused on very large systems near the border of quantum and classical physics \cite{boukobza2010nonlinear, boukobza2009phase}.
Here, we show that the results hold deep in the quantum limit.
It is interesting to note that chaos and entanglement are each exclusive to their respective classical and quantum domains and any connection is counterintuitive.
The correspondence is even more unexpected given that our system is so far from the classical limit \cite{gutzwiller1990chaos,berry1987bakerian}.

In Fig.\,\ref{fig:PhaseSpace}b, the entanglement entropy in the blue regions approaches 0.8, close to the maximum attainable value of 1.0 for a single qubit.
In Eq.\,\ref{Hamiltonian}, the Hamiltonian depends on time and, as a result, energy is not conserved.
Therefore, statistical mechanics would predict the values of observables using an ensemble with maximum entropy or, equivalently, an infinite temperature ensemble.
The observed density matrix approaching maximum entropy suggests that even small quantum systems undergoing unitary dynamics can appear to thermalize \cite{santos2012weak,polkovnikovmicroscopic,rigol2008thermalization}.

In the supplement, we numerically compute the evolution for larger systems and show that fluctuations decrease with increasing system size, as expected for finite-size systems approaching thermal equilibrium.
Additionally, we compute the behavior at larger values of $\kappa$ and show that all initial states obtain near maximal entropy, as opposed to the mixed phase space shown in Fig.\,\ref{fig:PhaseSpace}.
This further supports the idea that what we see in the experiment is the onset of thermalization in a small quantum system.

\begin{figure}[t!]
\includegraphics{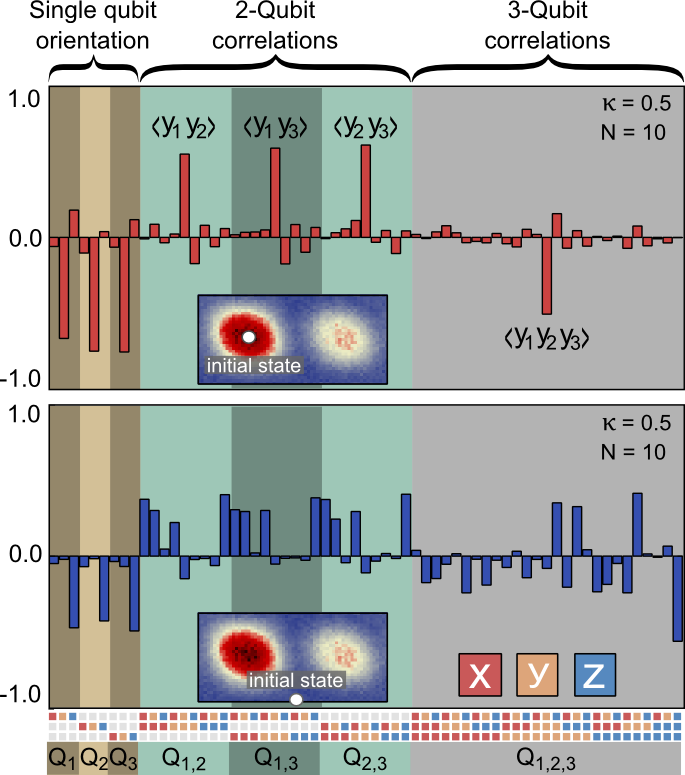} 
\caption {
\textbf{Multi-qubit entanglement.} We represent the three-qubit density-matrix for two initial states shown inset, one where the entropy was low (top) and one where the entropy was high (bottom). 
In both cases, the initial state was evolved for N = 10 time-steps and $\kappa = 0.5$.
Each bar indicates the expectation value of one possible combination of Pauli operators on the three qubits, the corresponding operator is shown using colored squares.  
The increase in multi-qubit correlations in the lower panel signifies that the contrast between high and low entropy is the result of entanglement.
}
\label{fig:Entanglement} 
\end{figure}

The observed single-qubit entropy can originate from two sources: entanglement with the other qubits and entanglement with the environment (decoherence).
In Fig.\,\ref{fig:Entanglement} we show that the contrast between high and low entropy results from entanglement amongst the qubits, confirming our assumption that the system is well isolated.
In order to distinguish these two effects, we measure the three-qubit density matrix.
Using these measurements, we compute the expectation values of all combinations of Pauli operators.
The first nine columns in  Fig.\,\ref{fig:Entanglement} contain operators only on a single qubit and thus provide information about local properties.
The remaining columns contain products of two- and three-qubit operators and describe correlations between the qubits.

In the top panel we consider an initial state whose entropy has increased by the least amount (most red), shown inset.
After ten time-steps, we see that each qubit is oriented along the $y$-axis as indicated by the first three peaks.
The qubits pointing along the same direction leads to classical correlations, as indicated by the remaining peaks among the two- and three-qubit correlations.
In the lower panel we consider an initial state whose entropy has increased by the largest amount (most blue).
In addition to the qubit orientations and classical correlations, we also find many significant peaks among the multi-qubit correlations.
These non-classical correlations are clear signatures of entanglement amongst the qubits.
Additionally, we find that the three-qubit state purity, a measure of decoherence, is equal for both of these states, showing that the contrast between high and low entropy is entirely the result of inter-qubit entanglement (see supplement).

\begin{figure}[t!]
\includegraphics{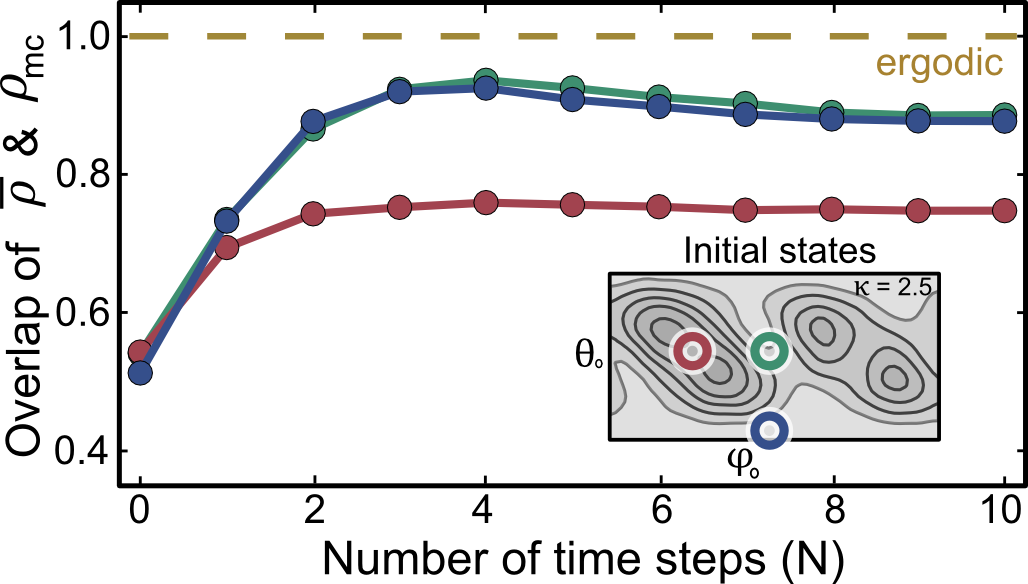} 
\caption {
\textbf{Ergodic dynamics.} The overlap of the time-averaged three-qubit density matrix with a microcanonical ensemble (see Eq.\,\ref{Overlap}) versus number of time steps $N$, for $\kappa = 2.5$.  We choose three different initial states, shown inset.  A value of 1.0 indicates that the dynamics are fully ergodic.
}
\label{fig:Thermalization} 
\end{figure}

%\psection{Figure 3.}
The advantage of studying statistical mechanics in a small quantum system is that we can directly check for ergodic motion in the three-qubit dynamics.
Using measurements of the full multi-qubit density matrix, we investigate the connection between ergodic dynamics in the full system and entropy production in subsystems.
Note that the full system is ideally in a pure state whose entropy is zero and stays zero as the system evolves - this is in stark contrast to subsystems which gain entropy over time through entanglement.
While the full system cannot thermalize in the sense of reaching maximum entropy, it can undergo ergodic motion (time averages being equal to state-space averages).
In statistical mechanics, a uniform average over states is given by the microcanonical ensemble.
In Fig.\,\ref{fig:Thermalization} we plot the overlap of the time-averaged density matrix $\bar{\rho}$ with a microcanonical ensemble $\rho_{\eqtext{mc}}$, given by
\begin{equation}\label{Overlap}
\eqtext{Overlap} = \Tr{\sqrt{\sqrt{\rho_{\eqtext{mc}}}\,\bar\rho\,\sqrt{\rho_{\eqtext{mc}}}}}
\end{equation}
The overlap of these two distributions approaching 1 would imply that time averages are equivalent to state-space averages for all measurable quantities.

We choose three different initial states: two are chosen from regions where subsystems had high entropy (blue \& green) and one from a region that had low entropy (red).
After just three steps, initial states where subsystems had high entropy approach a microcanonical ensemble to within  $94\%$.
Initial states where subsystems had low entropy fail to approach a microcanonical ensemble.
The reduction in overlap at long times  results from the accumulation of control errors and decoherence.
The strong overlap between time averages and state-space averages demonstrates that the three-qubit dynamics are ergodic and further supports the statistical mechanics framework for understanding the entropy production in single qubits.

%\psection{Conclusion}
It is interesting to know the generality of our results as they could provide a generic framework for studying quantum dynamics.
Numerical results suggest that ergodic behavior breaks down only when the evolution is highly constrained by conservation laws, such systems are referred to as integrable and represent models that are fine tuned and consequently rare \cite{rigol2008thermalization}.
Our choice of Hamiltonian was motivated by the lack of conserved quantities where only the total spin is conserved - not even energy is conserved.
We believe that our simple and clear descriptions of thermalization merely lay the foundation upon which many fundamental questions in non-equilibrium thermodynamics can be experimentally investigated.

\textbf{Acknowledgments:} We acknowledge discussions with  M. Fisher, P. Jessen, V. Madhok, C. Nayak,  A. Pattanayak, T. Prosen, A. Rahmani,  and D. Weld. This work was supported by the NSF under grants DMR-0907039 and DMR-1029764, the AFOSR under FA9550-10-1- 0110, and the ODNI, IARPA, through ARO grant W911NF-10-1-0334. Devices were made at the UCSB Nanofab Facility, part of the NSF-funded NNIN, and the NanoStructures Cleanroom Facility.

\textbf{Author Contributions:} C.N., P.R., and Y.C. designed and fabricated the sample and co-wrote the manuscript. C.N., P.R., and M.F. designed the experiment. C.N. performed the experiment and analyzed the data. M.K. and A.P. provided theoretical assistance. All members of the UCSB team contributed to the experimental setup and to the manuscript preparation.

\bibliography{cites}

\begin{thebibliography}{33}%
\makeatletter
\providecommand \@ifxundefined [1]{%
 \@ifx{#1\undefined}
}%
\providecommand \@ifnum [1]{%
 \ifnum #1\expandafter \@firstoftwo
 \else \expandafter \@secondoftwo
 \fi
}%
\providecommand \@ifx [1]{%
 \ifx #1\expandafter \@firstoftwo
 \else \expandafter \@secondoftwo
 \fi
}%
\providecommand \natexlab [1]{#1}%
\providecommand \enquote  [1]{``#1''}%
\providecommand \bibnamefont  [1]{#1}%
\providecommand \bibfnamefont [1]{#1}%
\providecommand \citenamefont [1]{#1}%
\providecommand \href@noop [0]{\@secondoftwo}%
\providecommand \href [0]{\begingroup \@sanitize@url \@href}%
\providecommand \@href[1]{\@@startlink{#1}\@@href}%
\providecommand \@@href[1]{\endgroup#1\@@endlink}%
\providecommand \@sanitize@url [0]{\catcode `\\12\catcode `\$12\catcode
  `\&12\catcode `\#12\catcode `\^12\catcode `\_12\catcode `\%12\relax}%
\providecommand \@@startlink[1]{}%
\providecommand \@@endlink[0]{}%
\providecommand \url  [0]{\begingroup\@sanitize@url \@url }%
\providecommand \@url [1]{\endgroup\@href {#1}{\urlprefix }}%
\providecommand \urlprefix  [0]{URL }%
\providecommand \Eprint [0]{\href }%
\providecommand \doibase [0]{http://dx.doi.org/}%
\providecommand \selectlanguage [0]{\@gobble}%
\providecommand \bibinfo  [0]{\@secondoftwo}%
\providecommand \bibfield  [0]{\@secondoftwo}%
\providecommand \translation [1]{[#1]}%
\providecommand \BibitemOpen [0]{}%
\providecommand \bibitemStop [0]{}%
\providecommand \bibitemNoStop [0]{.\EOS\space}%
\providecommand \EOS [0]{\spacefactor3000\relax}%
\providecommand \BibitemShut  [1]{\csname bibitem#1\endcsname}%
\let\auto@bib@innerbib\@empty
%</preamble>
\bibitem [{\citenamefont {Deutsch}(1991)}]{deutsch1991quantum}%
  \BibitemOpen
  \bibfield  {author} {\bibinfo {author} {\bibfnamefont {J.M.}\ \bibnamefont
  {Deutsch}},\ }\bibfield  {title} {\enquote {\bibinfo {title} {Quantum
  statistical mechanics in a closed system},}\ }\href@noop {} {\bibfield
  {journal} {\bibinfo  {journal} {Phys. Rev. A}\ }\textbf {\bibinfo {volume}
  {43}},\ \bibinfo {pages} {2046} (\bibinfo {year} {1991})}\BibitemShut
  {NoStop}%
\bibitem [{\citenamefont {Srednicki}(1994)}]{srednicki1994chaos}%
  \BibitemOpen
  \bibfield  {author} {\bibinfo {author} {\bibfnamefont {M.}~\bibnamefont
  {Srednicki}},\ }\bibfield  {title} {\enquote {\bibinfo {title} {Chaos and
  quantum thermalization},}\ }\href@noop {} {\bibfield  {journal} {\bibinfo
  {journal} {Phys. Rev. E}\ }\textbf {\bibinfo {volume} {50}},\ \bibinfo
  {pages} {888} (\bibinfo {year} {1994})}\BibitemShut {NoStop}%
\bibitem [{\citenamefont {Rigol}\ \emph {et~al.}(2008)\citenamefont {Rigol},
  \citenamefont {Dunjko},\ and\ \citenamefont
  {Olshanii}}]{rigol2008thermalization}%
  \BibitemOpen
  \bibfield  {author} {\bibinfo {author} {\bibfnamefont {M.}~\bibnamefont
  {Rigol}}, \bibinfo {author} {\bibfnamefont {V.}~\bibnamefont {Dunjko}}, \
  and\ \bibinfo {author} {\bibfnamefont {M.}~\bibnamefont {Olshanii}},\
  }\bibfield  {title} {\enquote {\bibinfo {title} {Thermalization and its
  mechanism for generic isolated quantum systems},}\ }\href@noop {} {\bibfield
  {journal} {\bibinfo  {journal} {Nature}\ }\textbf {\bibinfo {volume} {452}},\
  \bibinfo {pages} {854--858} (\bibinfo {year} {2008})}\BibitemShut {NoStop}%
\bibitem [{\citenamefont {Polkovnikov}\ \emph {et~al.}(2011)\citenamefont
  {Polkovnikov}, \citenamefont {Sengupta}, \citenamefont {Silva},\ and\
  \citenamefont {Vengalattore}}]{polkovnikov2011}%
  \BibitemOpen
  \bibfield  {author} {\bibinfo {author} {\bibfnamefont {A.}~\bibnamefont
  {Polkovnikov}}, \bibinfo {author} {\bibfnamefont {K.}~\bibnamefont
  {Sengupta}}, \bibinfo {author} {\bibfnamefont {A.}~\bibnamefont {Silva}}, \
  and\ \bibinfo {author} {\bibfnamefont {M.}~\bibnamefont {Vengalattore}},\
  }\bibfield  {title} {\enquote {\bibinfo {title} {Colloquium: Nonequilibrium
  dynamics of closed interacting quantum systems},}\ }\href@noop {} {\bibfield
  {journal} {\bibinfo  {journal} {Rev. of Mod. Phys.}\ }\textbf {\bibinfo
  {volume} {83}},\ \bibinfo {pages} {863} (\bibinfo {year} {2011})}\BibitemShut
  {NoStop}%
\bibitem [{\citenamefont {Ott}(2002)}]{ott2002chaos}%
  \BibitemOpen
  \bibfield  {author} {\bibinfo {author} {\bibfnamefont {E.}~\bibnamefont
  {Ott}},\ }\href@noop {} {\emph {\bibinfo {title} {Chaos in dynamical
  systems}}}\ (\bibinfo  {publisher} {Cambridge university press},\ \bibinfo
  {year} {2002})\BibitemShut {NoStop}%
\bibitem [{\citenamefont {Gutzwiller}(1990)}]{gutzwiller1990chaos}%
  \BibitemOpen
  \bibfield  {author} {\bibinfo {author} {\bibfnamefont {M.}~\bibnamefont
  {Gutzwiller}},\ }\href@noop {} {\emph {\bibinfo {title} {Chaos in classical
  and quantum mechanics}}},\ Vol.~\bibinfo {volume} {1}\ (\bibinfo  {publisher}
  {Springer Science \& Business Media},\ \bibinfo {year} {1990})\BibitemShut
  {NoStop}%
\bibitem [{\citenamefont {Popescu}\ \emph {et~al.}(2006)\citenamefont
  {Popescu}, \citenamefont {Short},\ and\ \citenamefont
  {Winter}}]{popescu2006entanglement}%
  \BibitemOpen
  \bibfield  {author} {\bibinfo {author} {\bibfnamefont {S.}~\bibnamefont
  {Popescu}}, \bibinfo {author} {\bibfnamefont {A.}~\bibnamefont {Short}}, \
  and\ \bibinfo {author} {\bibfnamefont {A.}~\bibnamefont {Winter}},\
  }\bibfield  {title} {\enquote {\bibinfo {title} {Entanglement and the
  foundations of statistical mechanics},}\ }\href@noop {} {\bibfield  {journal}
  {\bibinfo  {journal} {Nature Physics}\ }\textbf {\bibinfo {volume} {2}},\
  \bibinfo {pages} {754--758} (\bibinfo {year} {2006})}\BibitemShut {NoStop}%
\bibitem [{\citenamefont {Linden}\ \emph {et~al.}(2009)\citenamefont {Linden},
  \citenamefont {Popescu}, \citenamefont {Short},\ and\ \citenamefont
  {Winter}}]{linden2009quantum}%
  \BibitemOpen
  \bibfield  {author} {\bibinfo {author} {\bibfnamefont {N.}~\bibnamefont
  {Linden}}, \bibinfo {author} {\bibfnamefont {S.}~\bibnamefont {Popescu}},
  \bibinfo {author} {\bibfnamefont {A.~J.}\ \bibnamefont {Short}}, \ and\
  \bibinfo {author} {\bibfnamefont {A.}~\bibnamefont {Winter}},\ }\bibfield
  {title} {\enquote {\bibinfo {title} {Quantum mechanical evolution towards
  thermal equilibrium},}\ }\href@noop {} {\bibfield  {journal} {\bibinfo
  {journal} {Phys. Rev. E}\ }\textbf {\bibinfo {volume} {79}},\ \bibinfo
  {pages} {061103} (\bibinfo {year} {2009})}\BibitemShut {NoStop}%
\bibitem [{\citenamefont {Rahul}\ and\ \citenamefont
  {Huse}(2015)}]{huse2014review}%
  \BibitemOpen
  \bibfield  {author} {\bibinfo {author} {\bibfnamefont {N.}~\bibnamefont
  {Rahul}}\ and\ \bibinfo {author} {\bibfnamefont {D.}~\bibnamefont {Huse}},\
  }\bibfield  {title} {\enquote {\bibinfo {title} {Many body localization and
  thermalization in quantum statistical mechanics},}\ }\href@noop {} {\bibfield
   {journal} {\bibinfo  {journal} {Ann. Rev. of Cond. Matt. Phys.}\ }\textbf
  {\bibinfo {volume} {6}},\ \bibinfo {pages} {15--38} (\bibinfo {year}
  {2015})}\BibitemShut {NoStop}%
\bibitem [{\citenamefont {Polkovnikov}(2011)}]{polkovnikovmicroscopic}%
  \BibitemOpen
  \bibfield  {author} {\bibinfo {author} {\bibfnamefont {A.}~\bibnamefont
  {Polkovnikov}},\ }\bibfield  {title} {\enquote {\bibinfo {title} {Microscopic
  diagonal entropy and its connection to basic thermodynamic relations},}\
  }\href@noop {} {\bibfield  {journal} {\bibinfo  {journal} {Ann. of Phys.}\
  }\textbf {\bibinfo {volume} {326}},\ \bibinfo {pages} {486--499} (\bibinfo
  {year} {2011})}\BibitemShut {NoStop}%
\bibitem [{\citenamefont {Ermann}\ and\ \citenamefont
  {Shepelyansky}(2013)}]{ermann2013quantum}%
  \BibitemOpen
  \bibfield  {author} {\bibinfo {author} {\bibfnamefont {L.}~\bibnamefont
  {Ermann}}\ and\ \bibinfo {author} {\bibfnamefont {D.}~\bibnamefont
  {Shepelyansky}},\ }\bibfield  {title} {\enquote {\bibinfo {title} {Quantum
  gibbs distribution from dynamical thermalization in classical nonlinear
  lattices},}\ }\href@noop {} {\bibfield  {journal} {\bibinfo  {journal} {New
  J. of Phys.}\ }\textbf {\bibinfo {volume} {15}},\ \bibinfo {pages} {123004}
  (\bibinfo {year} {2013})}\BibitemShut {NoStop}%
\bibitem [{\citenamefont {Kinoshita}\ \emph {et~al.}(2006)\citenamefont
  {Kinoshita}, \citenamefont {Wenger},\ and\ \citenamefont
  {Weiss}}]{kinoshita2006quantum}%
  \BibitemOpen
  \bibfield  {author} {\bibinfo {author} {\bibfnamefont {T.}~\bibnamefont
  {Kinoshita}}, \bibinfo {author} {\bibfnamefont {T.}~\bibnamefont {Wenger}}, \
  and\ \bibinfo {author} {\bibfnamefont {D.}~\bibnamefont {Weiss}},\ }\bibfield
   {title} {\enquote {\bibinfo {title} {A quantum newton's cradle},}\
  }\href@noop {} {\bibfield  {journal} {\bibinfo  {journal} {Nature}\ }\textbf
  {\bibinfo {volume} {440}},\ \bibinfo {pages} {900--903} (\bibinfo {year}
  {2006})}\BibitemShut {NoStop}%
\bibitem [{\citenamefont {Klaers}\ \emph {et~al.}(2010)\citenamefont {Klaers},
  \citenamefont {Vewinger},\ and\ \citenamefont
  {Weitz}}]{klaers2010thermalization}%
  \BibitemOpen
  \bibfield  {author} {\bibinfo {author} {\bibfnamefont {J.}~\bibnamefont
  {Klaers}}, \bibinfo {author} {\bibfnamefont {F.}~\bibnamefont {Vewinger}}, \
  and\ \bibinfo {author} {\bibfnamefont {M.}~\bibnamefont {Weitz}},\ }\bibfield
   {title} {\enquote {\bibinfo {title} {Thermalization of a two-dimensional
  photonic gas in a white wall photon box},}\ }\href@noop {} {\bibfield
  {journal} {\bibinfo  {journal} {Nature Physics}\ }\textbf {\bibinfo {volume}
  {6}},\ \bibinfo {pages} {512--515} (\bibinfo {year} {2010})}\BibitemShut
  {NoStop}%
\bibitem [{\citenamefont {Cheneau}\ \emph {et~al.}(2012)\citenamefont
  {Cheneau}, \citenamefont {Barmettler}, \citenamefont {Poletti}, \citenamefont
  {Endres}, \citenamefont {Schau{\ss}}, \citenamefont {Fukuhara}, \citenamefont
  {Gross}, \citenamefont {Bloch}, \citenamefont {Kollath},\ and\ \citenamefont
  {Kuhr}}]{cheneau2012light}%
  \BibitemOpen
  \bibfield  {author} {\bibinfo {author} {\bibfnamefont {M.}~\bibnamefont
  {Cheneau}}, \bibinfo {author} {\bibfnamefont {P.}~\bibnamefont {Barmettler}},
  \bibinfo {author} {\bibfnamefont {D.}~\bibnamefont {Poletti}}, \bibinfo
  {author} {\bibfnamefont {M.}~\bibnamefont {Endres}}, \bibinfo {author}
  {\bibfnamefont {P.}~\bibnamefont {Schau{\ss}}}, \bibinfo {author}
  {\bibfnamefont {T.}~\bibnamefont {Fukuhara}}, \bibinfo {author}
  {\bibfnamefont {C.}~\bibnamefont {Gross}}, \bibinfo {author} {\bibfnamefont
  {I.}~\bibnamefont {Bloch}}, \bibinfo {author} {\bibfnamefont
  {C.}~\bibnamefont {Kollath}}, \ and\ \bibinfo {author} {\bibfnamefont
  {S.}~\bibnamefont {Kuhr}},\ }\bibfield  {title} {\enquote {\bibinfo {title}
  {Light-cone-like spreading of correlations in a quantum many-body system},}\
  }\href@noop {} {\bibfield  {journal} {\bibinfo  {journal} {Nature}\ }\textbf
  {\bibinfo {volume} {481}},\ \bibinfo {pages} {484--487} (\bibinfo {year}
  {2012})}\BibitemShut {NoStop}%
\bibitem [{\citenamefont {Trotzky}\ \emph {et~al.}(2012)\citenamefont
  {Trotzky}, \citenamefont {Chen}, \citenamefont {Flesch}, \citenamefont
  {McCulloch}, \citenamefont {Schollw{\"o}ck}, \citenamefont {Eisert},\ and\
  \citenamefont {Bloch}}]{trotzky2012probing}%
  \BibitemOpen
  \bibfield  {author} {\bibinfo {author} {\bibfnamefont {S.}~\bibnamefont
  {Trotzky}}, \bibinfo {author} {\bibfnamefont {Y.}~\bibnamefont {Chen}},
  \bibinfo {author} {\bibfnamefont {A.}~\bibnamefont {Flesch}}, \bibinfo
  {author} {\bibfnamefont {I.}~\bibnamefont {McCulloch}}, \bibinfo {author}
  {\bibfnamefont {U.}~\bibnamefont {Schollw{\"o}ck}}, \bibinfo {author}
  {\bibfnamefont {J.}~\bibnamefont {Eisert}}, \ and\ \bibinfo {author}
  {\bibfnamefont {I.}~\bibnamefont {Bloch}},\ }\bibfield  {title} {\enquote
  {\bibinfo {title} {Probing the relaxation towards equilibrium in an isolated
  strongly correlated one-dimensional bose gas},}\ }\href@noop {} {\bibfield
  {journal} {\bibinfo  {journal} {Nature Physics}\ }\textbf {\bibinfo {volume}
  {8}},\ \bibinfo {pages} {325--330} (\bibinfo {year} {2012})}\BibitemShut
  {NoStop}%
\bibitem [{\citenamefont {Langen}\ \emph {et~al.}(2013)\citenamefont {Langen},
  \citenamefont {Geiger}, \citenamefont {Kuhnert}, \citenamefont {Rauer},\ and\
  \citenamefont {Schmiedmayer}}]{langen2013local}%
  \BibitemOpen
  \bibfield  {author} {\bibinfo {author} {\bibfnamefont {T.}~\bibnamefont
  {Langen}}, \bibinfo {author} {\bibfnamefont {R.}~\bibnamefont {Geiger}},
  \bibinfo {author} {\bibfnamefont {M.}~\bibnamefont {Kuhnert}}, \bibinfo
  {author} {\bibfnamefont {B.}~\bibnamefont {Rauer}}, \ and\ \bibinfo {author}
  {\bibfnamefont {J.}~\bibnamefont {Schmiedmayer}},\ }\bibfield  {title}
  {\enquote {\bibinfo {title} {Local emergence of thermal correlations in an
  isolated quantum many-body system},}\ }\href@noop {} {\bibfield  {journal}
  {\bibinfo  {journal} {Nature Physics}\ }\textbf {\bibinfo {volume} {9}},\
  \bibinfo {pages} {640--643} (\bibinfo {year} {2013})}\BibitemShut {NoStop}%
\bibitem [{\citenamefont {Richerme}\ \emph {et~al.}(2014)\citenamefont
  {Richerme}, \citenamefont {Gong}, \citenamefont {Lee}, \citenamefont {Senko},
  \citenamefont {Smith}, \citenamefont {Foss-Feig}, \citenamefont {Michalakis},
  \citenamefont {Gorshkov},\ and\ \citenamefont {Monroe}}]{richerme2014non}%
  \BibitemOpen
  \bibfield  {author} {\bibinfo {author} {\bibfnamefont {P.}~\bibnamefont
  {Richerme}}, \bibinfo {author} {\bibfnamefont {Z.}~\bibnamefont {Gong}},
  \bibinfo {author} {\bibfnamefont {A.}~\bibnamefont {Lee}}, \bibinfo {author}
  {\bibfnamefont {C.}~\bibnamefont {Senko}}, \bibinfo {author} {\bibfnamefont
  {J.}~\bibnamefont {Smith}}, \bibinfo {author} {\bibfnamefont
  {M.}~\bibnamefont {Foss-Feig}}, \bibinfo {author} {\bibfnamefont
  {S.}~\bibnamefont {Michalakis}}, \bibinfo {author} {\bibfnamefont
  {A.}~\bibnamefont {Gorshkov}}, \ and\ \bibinfo {author} {\bibfnamefont
  {C.}~\bibnamefont {Monroe}},\ }\bibfield  {title} {\enquote {\bibinfo {title}
  {Non-local propagation of correlations in quantum systems with long-range
  interactions},}\ }\href@noop {} {\bibfield  {journal} {\bibinfo  {journal}
  {Nature}\ }\textbf {\bibinfo {volume} {511}},\ \bibinfo {pages} {198--201}
  (\bibinfo {year} {2014})}\BibitemShut {NoStop}%
\bibitem [{\citenamefont {Schreiber}\ \emph {et~al.}(2015)\citenamefont
  {Schreiber}, \citenamefont {Hodgman}, \citenamefont {Bordia}, \citenamefont
  {L{\"u}schen}, \citenamefont {Fischer}, \citenamefont {Vosk}, \citenamefont
  {Altman}, \citenamefont {Schneider},\ and\ \citenamefont
  {Bloch}}]{schreiber2015observation}%
  \BibitemOpen
  \bibfield  {author} {\bibinfo {author} {\bibfnamefont {M.}~\bibnamefont
  {Schreiber}}, \bibinfo {author} {\bibfnamefont {S.}~\bibnamefont {Hodgman}},
  \bibinfo {author} {\bibfnamefont {P.}~\bibnamefont {Bordia}}, \bibinfo
  {author} {\bibfnamefont {H.}~\bibnamefont {L{\"u}schen}}, \bibinfo {author}
  {\bibfnamefont {M.}~\bibnamefont {Fischer}}, \bibinfo {author} {\bibfnamefont
  {R.}~\bibnamefont {Vosk}}, \bibinfo {author} {\bibfnamefont {E.}~\bibnamefont
  {Altman}}, \bibinfo {author} {\bibfnamefont {U.}~\bibnamefont {Schneider}}, \
  and\ \bibinfo {author} {\bibfnamefont {I.}~\bibnamefont {Bloch}},\ }\bibfield
   {title} {\enquote {\bibinfo {title} {Observation of many-body localization
  of interacting fermions in a quasi-random optical lattice},}\ }\href@noop {}
  {\bibfield  {journal} {\bibinfo  {journal} {Science}\ }\textbf {\bibinfo
  {volume} {349}},\ \bibinfo {pages} {842} (\bibinfo {year}
  {2015})}\BibitemShut {NoStop}%
\bibitem [{\citenamefont {Wang}\ \emph {et~al.}(2004)\citenamefont {Wang},
  \citenamefont {Ghose}, \citenamefont {Sanders},\ and\ \citenamefont
  {Hu}}]{wang2004entanglement}%
  \BibitemOpen
  \bibfield  {author} {\bibinfo {author} {\bibfnamefont {X.}~\bibnamefont
  {Wang}}, \bibinfo {author} {\bibfnamefont {S.}~\bibnamefont {Ghose}},
  \bibinfo {author} {\bibfnamefont {B.~C}\ \bibnamefont {Sanders}}, \ and\
  \bibinfo {author} {\bibfnamefont {B.}~\bibnamefont {Hu}},\ }\bibfield
  {title} {\enquote {\bibinfo {title} {Entanglement as a signature of quantum
  chaos},}\ }\href@noop {} {\bibfield  {journal} {\bibinfo  {journal} {Phys.
  Rev. E}\ }\textbf {\bibinfo {volume} {70}},\ \bibinfo {pages} {016217}
  (\bibinfo {year} {2004})}\BibitemShut {NoStop}%
\bibitem [{\citenamefont {Ghose}\ \emph {et~al.}(2008)\citenamefont {Ghose},
  \citenamefont {Stock}, \citenamefont {Jessen}, \citenamefont {Lal},\ and\
  \citenamefont {Silberfarb}}]{ghose2008chaos}%
  \BibitemOpen
  \bibfield  {author} {\bibinfo {author} {\bibfnamefont {S.}~\bibnamefont
  {Ghose}}, \bibinfo {author} {\bibfnamefont {R.}~\bibnamefont {Stock}},
  \bibinfo {author} {\bibfnamefont {P.}~\bibnamefont {Jessen}}, \bibinfo
  {author} {\bibfnamefont {R.}~\bibnamefont {Lal}}, \ and\ \bibinfo {author}
  {\bibfnamefont {A.}~\bibnamefont {Silberfarb}},\ }\bibfield  {title}
  {\enquote {\bibinfo {title} {Chaos, entanglement, and decoherence in the
  quantum kicked top},}\ }\href@noop {} {\bibfield  {journal} {\bibinfo
  {journal} {Phys. Rev. A}\ }\textbf {\bibinfo {volume} {78}},\ \bibinfo
  {pages} {042318} (\bibinfo {year} {2008})}\BibitemShut {NoStop}%
\bibitem [{\citenamefont {Lombardi}\ and\ \citenamefont
  {Matzkin}(2011)}]{lombardi2011entanglement}%
  \BibitemOpen
  \bibfield  {author} {\bibinfo {author} {\bibfnamefont {M.}~\bibnamefont
  {Lombardi}}\ and\ \bibinfo {author} {\bibfnamefont {A.}~\bibnamefont
  {Matzkin}},\ }\bibfield  {title} {\enquote {\bibinfo {title} {Entanglement
  and chaos in the kicked top},}\ }\href@noop {} {\bibfield  {journal}
  {\bibinfo  {journal} {Phys. Rev. E}\ }\textbf {\bibinfo {volume} {83}},\
  \bibinfo {pages} {016207} (\bibinfo {year} {2011})}\BibitemShut {NoStop}%
\bibitem [{\citenamefont {Chaudhury}\ \emph {et~al.}(2009)\citenamefont
  {Chaudhury}, \citenamefont {Smith}, \citenamefont {Anderson}, \citenamefont
  {Ghose},\ and\ \citenamefont {Jessen}}]{chaudhury2009quantum}%
  \BibitemOpen
  \bibfield  {author} {\bibinfo {author} {\bibfnamefont {S.}~\bibnamefont
  {Chaudhury}}, \bibinfo {author} {\bibfnamefont {A.}~\bibnamefont {Smith}},
  \bibinfo {author} {\bibfnamefont {B.}~\bibnamefont {Anderson}}, \bibinfo
  {author} {\bibfnamefont {S.}~\bibnamefont {Ghose}}, \ and\ \bibinfo {author}
  {\bibfnamefont {P.}~\bibnamefont {Jessen}},\ }\bibfield  {title} {\enquote
  {\bibinfo {title} {Quantum signatures of chaos in a kicked top},}\
  }\href@noop {} {\bibfield  {journal} {\bibinfo  {journal} {Nature}\ }\textbf
  {\bibinfo {volume} {461}},\ \bibinfo {pages} {768--771} (\bibinfo {year}
  {2009})}\BibitemShut {NoStop}%
\bibitem [{\citenamefont {Haake}\ \emph {et~al.}(1987)\citenamefont {Haake},
  \citenamefont {Ku{\'s}},\ and\ \citenamefont {Scharf}}]{haake1987classical}%
  \BibitemOpen
  \bibfield  {author} {\bibinfo {author} {\bibfnamefont {F.}~\bibnamefont
  {Haake}}, \bibinfo {author} {\bibfnamefont {M.}~\bibnamefont {Ku{\'s}}}, \
  and\ \bibinfo {author} {\bibfnamefont {R.}~\bibnamefont {Scharf}},\
  }\bibfield  {title} {\enquote {\bibinfo {title} {Classical and quantum chaos
  for a kicked top},}\ }\href@noop {} {\bibfield  {journal} {\bibinfo
  {journal} {Zeitschrift f{\"u}r Physik B Condensed Matter}\ }\textbf {\bibinfo
  {volume} {65}},\ \bibinfo {pages} {381--395} (\bibinfo {year}
  {1987})}\BibitemShut {NoStop}%
\bibitem [{\citenamefont {Barends}\ \emph {et~al.}(2013)\citenamefont
  {Barends}, \citenamefont {Kelly}, \citenamefont {Megrant}, \citenamefont
  {Sank}, \citenamefont {Jeffrey}, \citenamefont {Chen}, \citenamefont {Yin},
  \citenamefont {Chiaro}, \citenamefont {Mutus}, \citenamefont {Neill} \emph
  {et~al.}}]{barends2013coherent}%
  \BibitemOpen
  \bibfield  {author} {\bibinfo {author} {\bibfnamefont {R.}~\bibnamefont
  {Barends}}, \bibinfo {author} {\bibfnamefont {J.}~\bibnamefont {Kelly}},
  \bibinfo {author} {\bibfnamefont {A.}~\bibnamefont {Megrant}}, \bibinfo
  {author} {\bibfnamefont {D.}~\bibnamefont {Sank}}, \bibinfo {author}
  {\bibfnamefont {E.}~\bibnamefont {Jeffrey}}, \bibinfo {author} {\bibfnamefont
  {Yu}~\bibnamefont {Chen}}, \bibinfo {author} {\bibfnamefont {Y.}~\bibnamefont
  {Yin}}, \bibinfo {author} {\bibfnamefont {B.}~\bibnamefont {Chiaro}},
  \bibinfo {author} {\bibfnamefont {J.}~\bibnamefont {Mutus}}, \bibinfo
  {author} {\bibfnamefont {C.}~\bibnamefont {Neill}},  \emph {et~al.},\
  }\bibfield  {title} {\enquote {\bibinfo {title} {Coherent josephson qubit
  suitable for scalable quantum integrated circuits},}\ }\href@noop {}
  {\bibfield  {journal} {\bibinfo  {journal} {Phys. Rev. Lett.}\ }\textbf
  {\bibinfo {volume} {111}},\ \bibinfo {pages} {080502} (\bibinfo {year}
  {2013})}\BibitemShut {NoStop}%
\bibitem [{\citenamefont {Chen}\ \emph {et~al.}(2014)\citenamefont {Chen},
  \citenamefont {Neill}, \citenamefont {Roushan}, \citenamefont {Leung},
  \citenamefont {Fang}, \citenamefont {Barends}, \citenamefont {Kelly},
  \citenamefont {Campbell}, \citenamefont {Chen}, \citenamefont {Chiaro} \emph
  {et~al.}}]{chen2014qubit}%
  \BibitemOpen
  \bibfield  {author} {\bibinfo {author} {\bibfnamefont {Yu}~\bibnamefont
  {Chen}}, \bibinfo {author} {\bibfnamefont {C.}~\bibnamefont {Neill}},
  \bibinfo {author} {\bibfnamefont {P.}~\bibnamefont {Roushan}}, \bibinfo
  {author} {\bibfnamefont {N.}~\bibnamefont {Leung}}, \bibinfo {author}
  {\bibfnamefont {M.}~\bibnamefont {Fang}}, \bibinfo {author} {\bibfnamefont
  {R.}~\bibnamefont {Barends}}, \bibinfo {author} {\bibfnamefont
  {J.}~\bibnamefont {Kelly}}, \bibinfo {author} {\bibfnamefont
  {B.}~\bibnamefont {Campbell}}, \bibinfo {author} {\bibfnamefont
  {Z.}~\bibnamefont {Chen}}, \bibinfo {author} {\bibfnamefont {B.}~\bibnamefont
  {Chiaro}},  \emph {et~al.},\ }\bibfield  {title} {\enquote {\bibinfo {title}
  {Qubit architecture with high coherence and fast tunable coupling},}\
  }\href@noop {} {\bibfield  {journal} {\bibinfo  {journal} {Phys. Rev. Lett.}\
  }\textbf {\bibinfo {volume} {113}},\ \bibinfo {pages} {220502} (\bibinfo
  {year} {2014})}\BibitemShut {NoStop}%
\bibitem [{\citenamefont {Geller}\ \emph {et~al.}(2015)\citenamefont {Geller},
  \citenamefont {Donate}, \citenamefont {Chen}, \citenamefont {Neill},
  \citenamefont {Roushan},\ and\ \citenamefont {Martinis}}]{geller2014tunable}%
  \BibitemOpen
  \bibfield  {author} {\bibinfo {author} {\bibfnamefont {M.}~\bibnamefont
  {Geller}}, \bibinfo {author} {\bibfnamefont {E.}~\bibnamefont {Donate}},
  \bibinfo {author} {\bibfnamefont {Y.}~\bibnamefont {Chen}}, \bibinfo {author}
  {\bibfnamefont {C.}~\bibnamefont {Neill}}, \bibinfo {author} {\bibfnamefont
  {P.}~\bibnamefont {Roushan}}, \ and\ \bibinfo {author} {\bibfnamefont
  {J.}~\bibnamefont {Martinis}},\ }\bibfield  {title} {\enquote {\bibinfo
  {title} {Tunable coupler for superconducting xmon qubits: Perturbative
  nonlinear model},}\ }\href@noop {} {\bibfield  {journal} {\bibinfo  {journal}
  {Phys. Rev. A}\ }\textbf {\bibinfo {volume} {92}},\ \bibinfo {pages} {012320}
  (\bibinfo {year} {2015})}\BibitemShut {NoStop}%
\bibitem [{\citenamefont {Neeley}\ \emph {et~al.}(2010)\citenamefont {Neeley},
  \citenamefont {Bialczak}, \citenamefont {Lenander}, \citenamefont {Lucero},
  \citenamefont {Mariantoni}, \citenamefont {O'Connell}, \citenamefont {Sank},
  \citenamefont {Wang}, \citenamefont {Weides}, \citenamefont {Wenner} \emph
  {et~al.}}]{neeley2010generation}%
  \BibitemOpen
  \bibfield  {author} {\bibinfo {author} {\bibfnamefont {M.}~\bibnamefont
  {Neeley}}, \bibinfo {author} {\bibfnamefont {R.~C}\ \bibnamefont {Bialczak}},
  \bibinfo {author} {\bibfnamefont {M.}~\bibnamefont {Lenander}}, \bibinfo
  {author} {\bibfnamefont {E.}~\bibnamefont {Lucero}}, \bibinfo {author}
  {\bibfnamefont {M.}~\bibnamefont {Mariantoni}}, \bibinfo {author}
  {\bibfnamefont {A.}~\bibnamefont {O'Connell}}, \bibinfo {author}
  {\bibfnamefont {D.}~\bibnamefont {Sank}}, \bibinfo {author} {\bibfnamefont
  {H.}~\bibnamefont {Wang}}, \bibinfo {author} {\bibfnamefont {M.}~\bibnamefont
  {Weides}}, \bibinfo {author} {\bibfnamefont {J.}~\bibnamefont {Wenner}},
  \emph {et~al.},\ }\bibfield  {title} {\enquote {\bibinfo {title} {Generation
  of three-qubit entangled states using superconducting phase qubits},}\
  }\href@noop {} {\bibfield  {journal} {\bibinfo  {journal} {Nature}\ }\textbf
  {\bibinfo {volume} {467}},\ \bibinfo {pages} {570--573} (\bibinfo {year}
  {2010})}\BibitemShut {NoStop}%
\bibitem [{\citenamefont {Khripkov}\ \emph {et~al.}(2013)\citenamefont
  {Khripkov}, \citenamefont {Cohen},\ and\ \citenamefont
  {Vardi}}]{khripkov2013coherence}%
  \BibitemOpen
  \bibfield  {author} {\bibinfo {author} {\bibfnamefont {C.}~\bibnamefont
  {Khripkov}}, \bibinfo {author} {\bibfnamefont {D.}~\bibnamefont {Cohen}}, \
  and\ \bibinfo {author} {\bibfnamefont {A.}~\bibnamefont {Vardi}},\ }\bibfield
   {title} {\enquote {\bibinfo {title} {Coherence dynamics of kicked
  bose-hubbard dimers: Interferometric signatures of chaos},}\ }\href@noop {}
  {\bibfield  {journal} {\bibinfo  {journal} {Phys. Rev. E}\ }\textbf {\bibinfo
  {volume} {87}},\ \bibinfo {pages} {012910} (\bibinfo {year}
  {2013})}\BibitemShut {NoStop}%
\bibitem [{\citenamefont {Madhok}\ \emph {et~al.}(2015)\citenamefont {Madhok},
  \citenamefont {Gupta}, \citenamefont {Hamel},\ and\ \citenamefont
  {Ghose}}]{madhok2013signatures}%
  \BibitemOpen
  \bibfield  {author} {\bibinfo {author} {\bibfnamefont {V.}~\bibnamefont
  {Madhok}}, \bibinfo {author} {\bibfnamefont {V.}~\bibnamefont {Gupta}},
  \bibinfo {author} {\bibfnamefont {A.}~\bibnamefont {Hamel}}, \ and\ \bibinfo
  {author} {\bibfnamefont {S.}~\bibnamefont {Ghose}},\ }\bibfield  {title}
  {\enquote {\bibinfo {title} {Signatures of chaos in the dynamics of quantum
  discord},}\ }\href@noop {} {\bibfield  {journal} {\bibinfo  {journal} {Phys.
  Rev. E}\ }\textbf {\bibinfo {volume} {91}},\ \bibinfo {pages} {032906}
  (\bibinfo {year} {2015})}\BibitemShut {NoStop}%
\bibitem [{\citenamefont {Boukobza}\ \emph {et~al.}(2010)\citenamefont
  {Boukobza}, \citenamefont {Moore}, \citenamefont {Cohen},\ and\ \citenamefont
  {Vardi}}]{boukobza2010nonlinear}%
  \BibitemOpen
  \bibfield  {author} {\bibinfo {author} {\bibfnamefont {E.}~\bibnamefont
  {Boukobza}}, \bibinfo {author} {\bibfnamefont {M.}~\bibnamefont {Moore}},
  \bibinfo {author} {\bibfnamefont {D.}~\bibnamefont {Cohen}}, \ and\ \bibinfo
  {author} {\bibfnamefont {A.}~\bibnamefont {Vardi}},\ }\bibfield  {title}
  {\enquote {\bibinfo {title} {Nonlinear phase-dynamics in a driven bosonic
  josephson junction},}\ }\href@noop {} {\bibfield  {journal} {\bibinfo
  {journal} {Phys. Rev. E}\ }\textbf {\bibinfo {volume} {104}},\ \bibinfo
  {pages} {240402} (\bibinfo {year} {2010})}\BibitemShut {NoStop}%
\bibitem [{\citenamefont {Boukobza}\ \emph {et~al.}(2009)\citenamefont
  {Boukobza}, \citenamefont {Chuchem}, \citenamefont {Cohen},\ and\
  \citenamefont {Vardi}}]{boukobza2009phase}%
  \BibitemOpen
  \bibfield  {author} {\bibinfo {author} {\bibfnamefont {E.}~\bibnamefont
  {Boukobza}}, \bibinfo {author} {\bibfnamefont {M.}~\bibnamefont {Chuchem}},
  \bibinfo {author} {\bibfnamefont {D.}~\bibnamefont {Cohen}}, \ and\ \bibinfo
  {author} {\bibfnamefont {A.}~\bibnamefont {Vardi}},\ }\bibfield  {title}
  {\enquote {\bibinfo {title} {Phase-diffusion dynamics in weakly coupled
  bose-einstein condensates},}\ }\href@noop {} {\bibfield  {journal} {\bibinfo
  {journal} {Phys. Rev. Lett.}\ }\textbf {\bibinfo {volume} {102}},\ \bibinfo
  {pages} {180403} (\bibinfo {year} {2009})}\BibitemShut {NoStop}%
\bibitem [{\citenamefont {Berry}(1987)}]{berry1987bakerian}%
  \BibitemOpen
  \bibfield  {author} {\bibinfo {author} {\bibfnamefont {M.}~\bibnamefont
  {Berry}},\ }\bibfield  {title} {\enquote {\bibinfo {title} {The bakerian
  lecture, 1987: quantum chaology},}\ }\href@noop {} {\bibfield  {journal}
  {\bibinfo  {journal} {Proceedings of the Royal Society of London}\ }\textbf
  {\bibinfo {volume} {413}},\ \bibinfo {pages} {183--198} (\bibinfo {year}
  {1987})}\BibitemShut {NoStop}%
\bibitem [{\citenamefont {Santos}\ \emph {et~al.}(2012)\citenamefont {Santos},
  \citenamefont {Polkovnikov},\ and\ \citenamefont {Rigol}}]{santos2012weak}%
  \BibitemOpen
  \bibfield  {author} {\bibinfo {author} {\bibfnamefont {L.}~\bibnamefont
  {Santos}}, \bibinfo {author} {\bibfnamefont {A.}~\bibnamefont {Polkovnikov}},
  \ and\ \bibinfo {author} {\bibfnamefont {M.}~\bibnamefont {Rigol}},\
  }\bibfield  {title} {\enquote {\bibinfo {title} {Weak and strong typicality
  in quantum systems},}\ }\href@noop {} {\bibfield  {journal} {\bibinfo
  {journal} {Phys. Rev. E}\ }\textbf {\bibinfo {volume} {86}},\ \bibinfo
  {pages} {010102} (\bibinfo {year} {2012})}\BibitemShut {NoStop}%
\end{thebibliography}%


\begin{thebibliography}{7}%
\makeatletter
\providecommand \@ifxundefined [1]{%
 \@ifx{#1\undefined}
}%
\providecommand \@ifnum [1]{%
 \ifnum #1\expandafter \@firstoftwo
 \else \expandafter \@secondoftwo
 \fi
}%
\providecommand \@ifx [1]{%
 \ifx #1\expandafter \@firstoftwo
 \else \expandafter \@secondoftwo
 \fi
}%
\providecommand \natexlab [1]{#1}%
\providecommand \enquote  [1]{``#1''}%
\providecommand \bibnamefont  [1]{#1}%
\providecommand \bibfnamefont [1]{#1}%
\providecommand \citenamefont [1]{#1}%
\providecommand \href@noop [0]{\@secondoftwo}%
\providecommand \href [0]{\begingroup \@sanitize@url \@href}%
\providecommand \@href[1]{\@@startlink{#1}\@@href}%
\providecommand \@@href[1]{\endgroup#1\@@endlink}%
\providecommand \@sanitize@url [0]{\catcode `\\12\catcode `\$12\catcode
  `\&12\catcode `\#12\catcode `\^12\catcode `\_12\catcode `\%12\relax}%
\providecommand \@@startlink[1]{}%
\providecommand \@@endlink[0]{}%
\providecommand \url  [0]{\begingroup\@sanitize@url \@url }%
\providecommand \@url [1]{\endgroup\@href {#1}{\urlprefix }}%
\providecommand \urlprefix  [0]{URL }%
\providecommand \Eprint [0]{\href }%
\providecommand \doibase [0]{http://dx.doi.org/}%
\providecommand \selectlanguage [0]{\@gobble}%
\providecommand \bibinfo  [0]{\@secondoftwo}%
\providecommand \bibfield  [0]{\@secondoftwo}%
\providecommand \translation [1]{[#1]}%
\providecommand \BibitemOpen [0]{}%
\providecommand \bibitemStop [0]{}%
\providecommand \bibitemNoStop [0]{.\EOS\space}%
\providecommand \EOS [0]{\spacefactor3000\relax}%
\providecommand \BibitemShut  [1]{\csname bibitem#1\endcsname}%
\let\auto@bib@innerbib\@empty
%</preamble>
\bibitem [{\citenamefont {Chen}\ \emph
  {et~al.}(2014{\natexlab{a}})\citenamefont {Chen}, \citenamefont {Neill},
  \citenamefont {Roushan}, \citenamefont {Leung}, \citenamefont {Fang},
  \citenamefont {Barends}, \citenamefont {Kelly}, \citenamefont {Campbell},
  \citenamefont {Chen}, \citenamefont {Chiaro} \emph {et~al.}}]{chen2014qubit}%
  \BibitemOpen
  \bibfield  {author} {\bibinfo {author} {\bibfnamefont {Yu}~\bibnamefont
  {Chen}}, \bibinfo {author} {\bibfnamefont {C.}~\bibnamefont {Neill}},
  \bibinfo {author} {\bibfnamefont {P.}~\bibnamefont {Roushan}}, \bibinfo
  {author} {\bibfnamefont {N.}~\bibnamefont {Leung}}, \bibinfo {author}
  {\bibfnamefont {M.}~\bibnamefont {Fang}}, \bibinfo {author} {\bibfnamefont
  {R.}~\bibnamefont {Barends}}, \bibinfo {author} {\bibfnamefont
  {J.}~\bibnamefont {Kelly}}, \bibinfo {author} {\bibfnamefont
  {B.}~\bibnamefont {Campbell}}, \bibinfo {author} {\bibfnamefont
  {Z.}~\bibnamefont {Chen}}, \bibinfo {author} {\bibfnamefont {B.}~\bibnamefont
  {Chiaro}},  \emph {et~al.},\ }\bibfield  {title} {\enquote {\bibinfo {title}
  {Qubit architecture with high coherence and fast tunable coupling},}\
  }\href@noop {} {\bibfield  {journal} {\bibinfo  {journal} {Phys. Rev. Lett.}\
  }\textbf {\bibinfo {volume} {113}},\ \bibinfo {pages} {220502} (\bibinfo
  {year} {2014}{\natexlab{a}})}\BibitemShut {NoStop}%
\bibitem [{\citenamefont {Barends}\ \emph {et~al.}(2013)\citenamefont
  {Barends}, \citenamefont {Kelly}, \citenamefont {Megrant}, \citenamefont
  {Sank}, \citenamefont {Jeffrey}, \citenamefont {Chen}, \citenamefont {Yin},
  \citenamefont {Chiaro}, \citenamefont {Mutus}, \citenamefont {Neill} \emph
  {et~al.}}]{barends2013coherent}%
  \BibitemOpen
  \bibfield  {author} {\bibinfo {author} {\bibfnamefont {R.}~\bibnamefont
  {Barends}}, \bibinfo {author} {\bibfnamefont {J.}~\bibnamefont {Kelly}},
  \bibinfo {author} {\bibfnamefont {A.}~\bibnamefont {Megrant}}, \bibinfo
  {author} {\bibfnamefont {D.}~\bibnamefont {Sank}}, \bibinfo {author}
  {\bibfnamefont {E.}~\bibnamefont {Jeffrey}}, \bibinfo {author} {\bibfnamefont
  {Yu}~\bibnamefont {Chen}}, \bibinfo {author} {\bibfnamefont {Y.}~\bibnamefont
  {Yin}}, \bibinfo {author} {\bibfnamefont {B.}~\bibnamefont {Chiaro}},
  \bibinfo {author} {\bibfnamefont {J.}~\bibnamefont {Mutus}}, \bibinfo
  {author} {\bibfnamefont {C.}~\bibnamefont {Neill}},  \emph {et~al.},\
  }\bibfield  {title} {\enquote {\bibinfo {title} {Coherent josephson qubit
  suitable for scalable quantum integrated circuits},}\ }\href@noop {}
  {\bibfield  {journal} {\bibinfo  {journal} {Phys. Rev. Lett.}\ }\textbf
  {\bibinfo {volume} {111}},\ \bibinfo {pages} {080502} (\bibinfo {year}
  {2013})}\BibitemShut {NoStop}%
\bibitem [{\citenamefont {Barends}\ \emph {et~al.}(2014)\citenamefont
  {Barends}, \citenamefont {Kelly}, \citenamefont {Megrant}, \citenamefont
  {Veitia}, \citenamefont {Sank}, \citenamefont {Jeffrey}, \citenamefont
  {White}, \citenamefont {Mutus}, \citenamefont {Fowler}, \citenamefont
  {Campbell} \emph {et~al.}}]{barends2014superconducting}%
  \BibitemOpen
  \bibfield  {author} {\bibinfo {author} {\bibfnamefont {R}~\bibnamefont
  {Barends}}, \bibinfo {author} {\bibfnamefont {J}~\bibnamefont {Kelly}},
  \bibinfo {author} {\bibfnamefont {A}~\bibnamefont {Megrant}}, \bibinfo
  {author} {\bibfnamefont {A}~\bibnamefont {Veitia}}, \bibinfo {author}
  {\bibfnamefont {D}~\bibnamefont {Sank}}, \bibinfo {author} {\bibfnamefont
  {E}~\bibnamefont {Jeffrey}}, \bibinfo {author} {\bibfnamefont
  {TC}~\bibnamefont {White}}, \bibinfo {author} {\bibfnamefont {J}~\bibnamefont
  {Mutus}}, \bibinfo {author} {\bibfnamefont {AG}~\bibnamefont {Fowler}},
  \bibinfo {author} {\bibfnamefont {B}~\bibnamefont {Campbell}},  \emph
  {et~al.},\ }\bibfield  {title} {\enquote {\bibinfo {title} {Superconducting
  quantum circuits at the surface code threshold for fault tolerance},}\
  }\href@noop {} {\bibfield  {journal} {\bibinfo  {journal} {Nature}\ }\textbf
  {\bibinfo {volume} {508}},\ \bibinfo {pages} {500--503} (\bibinfo {year}
  {2014})}\BibitemShut {NoStop}%
\bibitem [{\citenamefont {Koch}\ \emph {et~al.}(2007)\citenamefont {Koch},
  \citenamefont {Terri}, \citenamefont {Gambetta}, \citenamefont {Houck},
  \citenamefont {Schuster}, \citenamefont {Majer}, \citenamefont {Blais},
  \citenamefont {Devoret}, \citenamefont {Girvin},\ and\ \citenamefont
  {Schoelkopf}}]{koch2007charge}%
  \BibitemOpen
  \bibfield  {author} {\bibinfo {author} {\bibfnamefont {J.}~\bibnamefont
  {Koch}}, \bibinfo {author} {\bibfnamefont {Y.}~\bibnamefont {Terri}},
  \bibinfo {author} {\bibfnamefont {J.}~\bibnamefont {Gambetta}}, \bibinfo
  {author} {\bibfnamefont {A.}~\bibnamefont {Houck}}, \bibinfo {author}
  {\bibfnamefont {D.}~\bibnamefont {Schuster}}, \bibinfo {author}
  {\bibfnamefont {J.}~\bibnamefont {Majer}}, \bibinfo {author} {\bibfnamefont
  {A.}~\bibnamefont {Blais}}, \bibinfo {author} {\bibfnamefont
  {M.}~\bibnamefont {Devoret}}, \bibinfo {author} {\bibfnamefont
  {S.}~\bibnamefont {Girvin}}, \ and\ \bibinfo {author} {\bibfnamefont
  {R.}~\bibnamefont {Schoelkopf}},\ }\bibfield  {title} {\enquote {\bibinfo
  {title} {Charge-insensitive qubit design derived from the cooper pair box},}\
  }\href@noop {} {\bibfield  {journal} {\bibinfo  {journal} {Phys. Rev. A}\
  }\textbf {\bibinfo {volume} {76}},\ \bibinfo {pages} {042319} (\bibinfo
  {year} {2007})}\BibitemShut {NoStop}%
\bibitem [{\citenamefont {Chen}\ \emph
  {et~al.}(2014{\natexlab{b}})\citenamefont {Chen}, \citenamefont {Megrant},
  \citenamefont {Kelly}, \citenamefont {Barends}, \citenamefont {Bochmann},
  \citenamefont {Chen}, \citenamefont {Chiaro}, \citenamefont {Dunsworth},
  \citenamefont {Jeffrey}, \citenamefont {Mutus},\ and\ \citenamefont {et.
  al.}}]{chen2014fabrication}%
  \BibitemOpen
  \bibfield  {author} {\bibinfo {author} {\bibfnamefont {Z.}~\bibnamefont
  {Chen}}, \bibinfo {author} {\bibfnamefont {A.}~\bibnamefont {Megrant}},
  \bibinfo {author} {\bibfnamefont {J.}~\bibnamefont {Kelly}}, \bibinfo
  {author} {\bibfnamefont {R.}~\bibnamefont {Barends}}, \bibinfo {author}
  {\bibfnamefont {J.}~\bibnamefont {Bochmann}}, \bibinfo {author}
  {\bibfnamefont {Y.}~\bibnamefont {Chen}}, \bibinfo {author} {\bibfnamefont
  {B.}~\bibnamefont {Chiaro}}, \bibinfo {author} {\bibfnamefont
  {A.}~\bibnamefont {Dunsworth}}, \bibinfo {author} {\bibfnamefont
  {E.}~\bibnamefont {Jeffrey}}, \bibinfo {author} {\bibfnamefont
  {J.}~\bibnamefont {Mutus}}, \ and\ \bibinfo {author} {\bibnamefont {et.
  al.}},\ }\bibfield  {title} {\enquote {\bibinfo {title} {Fabrication and
  characterization of aluminum airbridges for superconducting microwave
  circuits},}\ }\href@noop {} {\bibfield  {journal} {\bibinfo  {journal} {App.
  Phys. Lett.}\ }\textbf {\bibinfo {volume} {104}},\ \bibinfo {pages} {052602}
  (\bibinfo {year} {2014}{\natexlab{b}})}\BibitemShut {NoStop}%
\bibitem [{\citenamefont {Chow}(2010)}]{chow2010thesis}%
  \BibitemOpen
  \bibfield  {author} {\bibinfo {author} {\bibfnamefont {J.}~\bibnamefont
  {Chow}},\ }\bibfield  {title} {\enquote {\bibinfo {title} {Quantum
  information processing with superconducting qubits},}\ }\href@noop {}
  {\bibfield  {journal} {\bibinfo  {journal} {PhD dissertation, Yale}\ }
  (\bibinfo {year} {2010})}\BibitemShut {NoStop}%
\bibitem [{\citenamefont {Neeley}\ \emph {et~al.}(2010)\citenamefont {Neeley},
  \citenamefont {Bialczak}, \citenamefont {Lenander}, \citenamefont {Lucero},
  \citenamefont {Mariantoni}, \citenamefont {O'Connell}, \citenamefont {Sank},
  \citenamefont {Wang}, \citenamefont {Weides}, \citenamefont {Wenner} \emph
  {et~al.}}]{neeley2010generation}%
  \BibitemOpen
  \bibfield  {author} {\bibinfo {author} {\bibfnamefont {M.}~\bibnamefont
  {Neeley}}, \bibinfo {author} {\bibfnamefont {R.~C}\ \bibnamefont {Bialczak}},
  \bibinfo {author} {\bibfnamefont {M.}~\bibnamefont {Lenander}}, \bibinfo
  {author} {\bibfnamefont {E.}~\bibnamefont {Lucero}}, \bibinfo {author}
  {\bibfnamefont {M.}~\bibnamefont {Mariantoni}}, \bibinfo {author}
  {\bibfnamefont {A.}~\bibnamefont {O'Connell}}, \bibinfo {author}
  {\bibfnamefont {D.}~\bibnamefont {Sank}}, \bibinfo {author} {\bibfnamefont
  {H.}~\bibnamefont {Wang}}, \bibinfo {author} {\bibfnamefont {M.}~\bibnamefont
  {Weides}}, \bibinfo {author} {\bibfnamefont {J.}~\bibnamefont {Wenner}},
  \emph {et~al.},\ }\bibfield  {title} {\enquote {\bibinfo {title} {Generation
  of three-qubit entangled states using superconducting phase qubits},}\
  }\href@noop {} {\bibfield  {journal} {\bibinfo  {journal} {Nature}\ }\textbf
  {\bibinfo {volume} {467}},\ \bibinfo {pages} {570--573} (\bibinfo {year}
  {2010})}\BibitemShut {NoStop}%
\end{thebibliography}%

\end{document}

\title{Supplementary Information for "Ergodic dynamics and thermalization in an isolated quantum system"}

\author{C. Neill$^{1}$}
\thanks{These authors contributed equally to this work.}
\author{P. Roushan$^{2}$}
\thanks{These authors contributed equally to this work.}
\author{M. Fang$^{1}$}
\thanks{These authors contributed equally to this work.}
\author{Y. Chen$^{2}$}
\thanks{These authors contributed equally to this work.}
\author{M. Kolodrubetz$^{3}$}

\author{Z. Chen$^{1}$}
\author{A. Megrant$^{2}$}
\author{R. Barends$^{2}$}
\author{B. Campbell$^{1}$}
\author{B. Chiaro$^{1}$}
\author{A. Dunsworth$^{1}$}
\author{E. Jeffrey$^{2}$}
\author{J. Kelly$^{2}$}
\author{J. Mutus$^{2}$}
\author{P. J. J. O'Malley$^{1}$}
\author{C. Quintana$^{1}$}
\author{D. Sank$^{2}$}
\author{A. Vainsencher$^{1}$}
\author{J. Wenner$^{1}$}
\author{T. C. White$^{2}$}
\author{A. Polkovnikov$^{3}$}
\author{J. M. Martinis$^{1,2}$}

\email{martinis@physics.ucsb.edu}

\affiliation{$^{1}$Department of Physics, University of California, Santa Barbara,
CA 93106-9530, USA}

\affiliation{$^{2}$Google Inc., Santa Barbara,
CA 93117, USA}

\affiliation{$^{3}$Department of Physics, Boston University, Boston,
MA 02215, USA}

\maketitle

\section{{I. Qubit architecture}}

There are two fundamental requirements for implementing the quantum dynamics demonstrated in this work:  a high level of individual control and long coherence times.
In pursuit of these goals, we have designed three transmon qubits with tunable qubit-qubit coupling, tunable frequencies and individual microwave control \cite{chen2014qubit}.
Transmon qubits, the Xmon design in particular, have been shown to have long coherence times \cite{barends2013coherent, barends2014superconducting, koch2007charge}.
The qubits are arranged into a ring in order to explore the model outlined in the main text beyond the more technologically straight-forward two-qubit realization.

\begin{figure}[t!]
\includegraphics{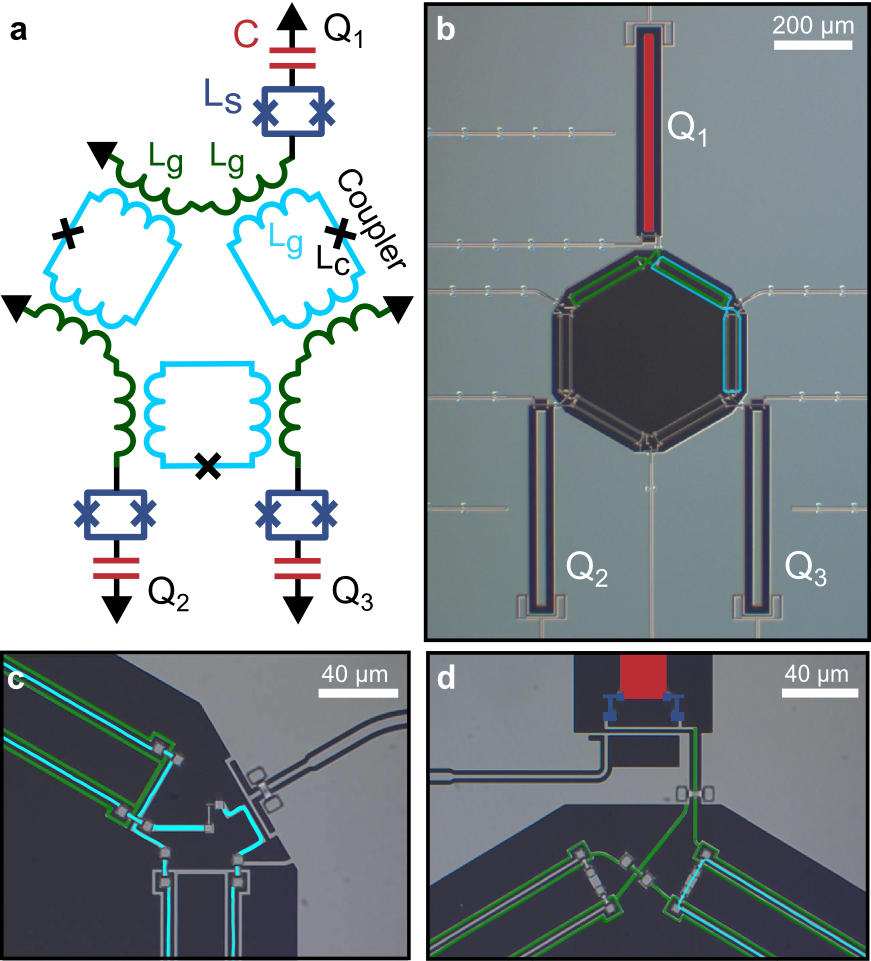} 
\caption {
\textbf{gmon architecture.}  We have designed a modified version of an Xmon qubit with tunable inter-qubit coupling. Panel \textbf{a} shows the circuit diagram for the device.  Each qubit is represented as a capacitor ($C = 75$\,fF) in series with a DC SQUID ($L_S = 8.1$\,nH) and two inductors ($L_g = 0.35$\,nH).  Each inductor is flux coupled to an RF SQUID (`coupler') through a mutual inductance ($M = 0.2$\,nH).  Applying a flux to the RF SQUID loop modulates the effective junction inductance ($L_C = 0.9$\,nH) and consequently the inter-qubit coupling strength. The effective SQUID inductances are the values at zero external flux. \textbf{b,}  Optical micrograph of the device.  Grey regions correspond to aluminum; black regions are where the aluminum has been etched away to expose the underlying sapphire substrate to define the qubits and wiring.  \textbf{c,d,}  Optical micrographs showing the coupler and qubit flux biases.  The qubit and coupler inductors $L_g$ can be seen highlighted in green and cyan, respectively.  All crossover connections are made using dielectric-free airbridges \cite{chen2014fabrication}.
}
\label{fig:Device} 
\end{figure}

A circuit diagram and optical micrographs of our gmon qubits are shown in Fig.\,\ref{fig:Device}.  
The individual qubits are composed of a capacitor (red), a DC SQUID (blue), and two inductors in series to ground (green).
The capacitor and SQUID form the basis of the standard Xmon qubit with the added inductors each allowing for tunable coupling to a neighboring qubit.

Tunable coupling is achieved through a mutual inductance to a loop containing a Josephson junction (cyan).
This loop mediates the interaction between pairs of qubits.
An excitation in either qubit generates a current in this loop which then excites the neighboring qubit.
The strength of the qubit-qubit interaction $g$ is modulated by applying a flux to the coupler loop; this flux sets the effective junction inductance.
If the junction inductance is large, then a smaller current will flow through the coupler loop and the coupling will become weaker.
For this device, the interaction strengths $g$/2$\pi$ were tunable from +5 MHz to -15 MHz; a value of -5 MHz was used for all of the experiments.

\begin{figure}[t!]
\includegraphics{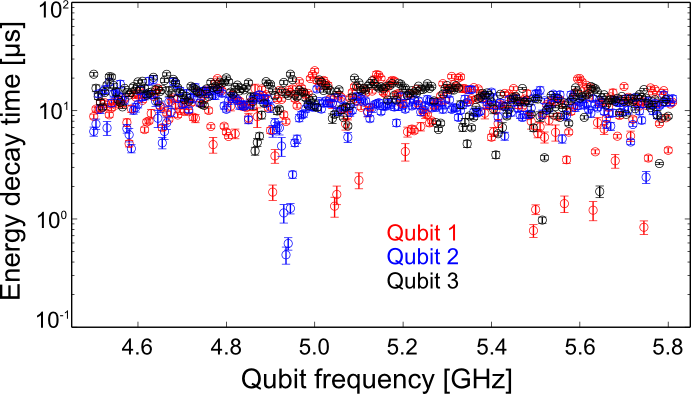} 
\caption {
\textbf{Energy decay time $T_1$.}  The energy decay time of each qubit as a function of the qubit frequency.  Each data point is measured by exciting the qubit, detuning it to the desired frequency,  waiting a variable delay time, measuring the qubit excited state probability and fitting the decay curve to an exponential.  The experimental results in this work were obtained near 5.7\,GHz where the decay times ranged from 12 to 18 $\mu$s.
}
\label{fig:Coherence} 
\end{figure}

The energy decay times $T_1$ for all three qubits are shown in Fig.\,\ref{fig:Coherence} versus qubit frequency.
During the thermalization experiments, the qubits were operated near 5.7\,GHz where the decay times of the three qubits were between 12 and 18\,$\mu$s.
Each experimental sequence ran for at most 500\,ns, excluding measurement.
The time scales of the experiment were an order of magnitude below the energy decay times.
The single-qubit dephasing times measured with Ramsey, however, ranged between 2 to 4\,$\mu$s, closer to the relevant time scales of the experiment.
In the single qubit experiments shown in Fig\,2 of the main text, decoherence is indistinguishable from entanglement with the other qubits. 
Measurements of the full three-qubit density matrix, however, allow us to separate decoherence from entanglement through multi-qubit correlation functions \cite{chow2010thesis}.

\section{II. Pulse sequence}

\begin{figure}[t!]
\includegraphics{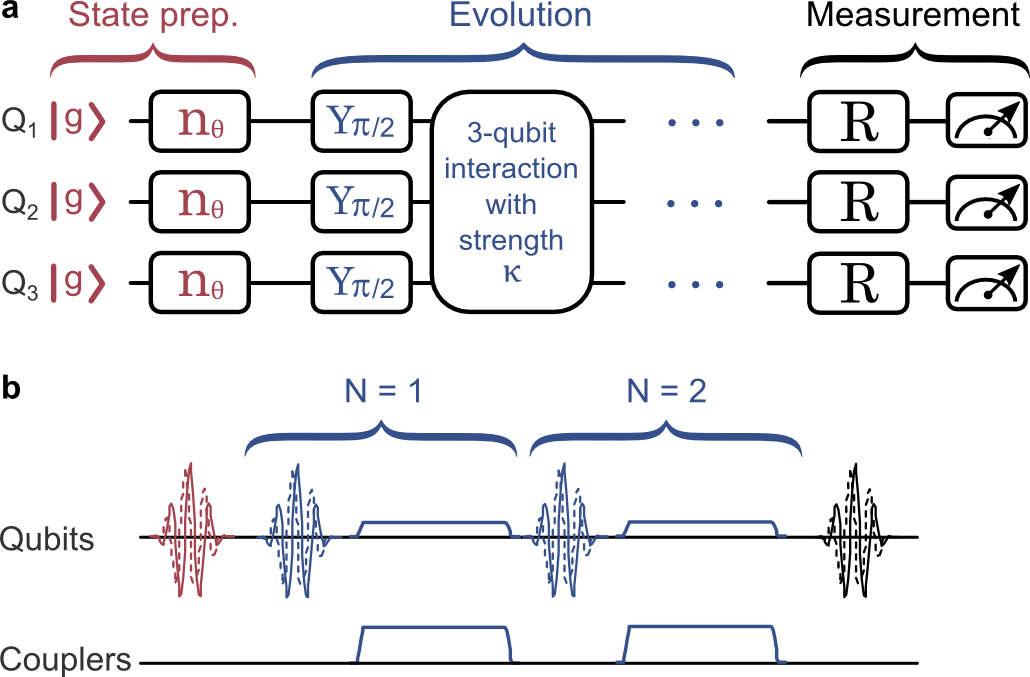} 
\caption {
\textbf{Pulse sequence and control waveforms. a,} Gate sequence used to study ergodic dynamics and thermalization.  First, each qubit is prepared in the ground state by waiting several energy decay times.  Next, we  rotate each qubit into the state $\ket{\theta_0, \phi_0}$ through a rotation around the axis $n = -\sin(\phi_0)\,\hat{x} + \cos(\phi_0)\,\hat{y}$ by angle $\theta_0$. This initial state is then evolved by $N$ applications of a rotation around the $y$-axis by $\pi/2$ and a symmetric multi-qubit interaction.   Following the evolution, the density matrix of either individual qubits or of the full system is determined using state tomography.  State tomography consists of a rotation followed by a measurement along the z-axis.  This is repeated for different rotation angles and axes to reconstruct the density matrix.
\textbf{b.} The control waveforms used to implement the gate sequence are shown for $N = 2$.  Oscillatory signals correspond to resonant microwave pulses used to rotate the single-qubit states.  The amplitude and phase of the control waveform determine the rotation angle and axis respectively.  Square pulses applied to the coupler and qubit SQUID loops are used to turn on the multi-qubit interaction and to maintain the qubits on resonance.
}
\label{fig:PulseSequence} 
\end{figure}

In Fig.\,\ref{fig:PulseSequence} we show the pulse sequence and corresponding control waveforms used to implement the experiments in the main text.
The pulse sequence can be broken up into three sections: state preparation, evolution and measurement.
The initial states $\ket{\theta_0, \phi_0}$ were prepared in 40\,ns using resonant microwave pulses, shown as a red oscillatory signal in the lower panel.
The amplitude and length of the microwave pulse set the angle $\theta_0$; the phase of the microwave pulse sets $\phi_0$.
Each time step in the evolution then consists of two parts: a $y$-rotation and a symmetric three-qubit interaction.
The $y$-rotation is achieved in 20\,ns using a resonant microwave pulse shown in blue.
The three-qubit interaction is performed by applying a square pulse to each coupling circuit, the duration of which sets $\kappa$.
During the interaction, square pulses are used to maintain the qubits on resonance with one another as the coupler pulses cause the qubits to shift in frequency.
We additionally calibrate for cross-talk between the six low-frequency control lines:  three lines which tune the qubit frequencies and three which tune the coupling.
The cross-talk matrix $dV$ defined as $V_{\eqtext{actual}} = \left( 1 + dV\right)V_{\eqtext{ideal}}$ was measured to be 
\begin{equation}
dV = \bordermatrix{~ & \eqtext{cp}12 & \eqtext{cp}23 & \eqtext{cp}31 & \eqtext{q}1 & \eqtext{q}2 & \eqtext{q}3 \cr
& 0.00 & 0.09 &  0.07 & -0.08 & -0.05 &  0.15 \cr
& 0.03 & 0.00 &  0.05 &  0.14 &  0.06 & -0.07 \cr
& 0.09 & 0.11 &  0.00 & -0.35 &  0.15 & -0.04 \cr
& 0.04 & 0.00 & -0.05 &  0.00 &  0.05 & -0.04 \cr
& -0.02 & 0.02 &  0.02 &  0.01 &  0.00 &  0.03 \cr
& 0.02 & 0.02 & -0.02 & -0.01 &  0.04 &  0.00 }
\nonumber
\end{equation}

\begin{figure}[t!]
\includegraphics{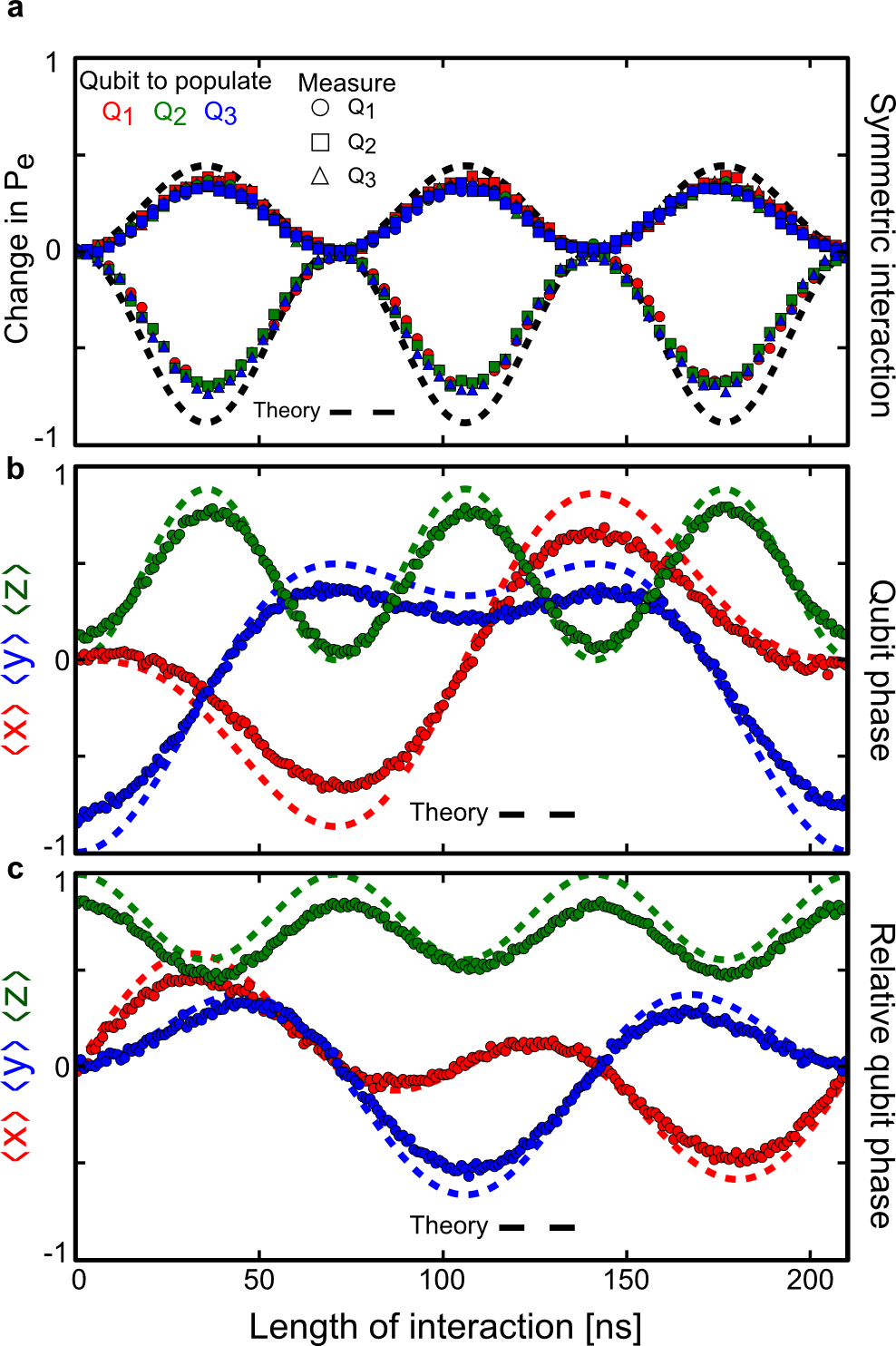} 
\caption {
\textbf{Characterizing the 3-qubit interaction. a,} Here we demonstrate that the inter-qubit interaction energies are all of equal strength and that the qubits are on resonance during the interaction.  This is done by exciting a single qubit, turning on the interaction for a variable length of time (horizontal axis) and then measuring all three qubit excited state probabilities $P_e$.  We plot the change in $P_e$ relative to having waited the corresponding length of time.  We then repeat the experiment exciting different qubits, resulting in a total of 9 curves.  The symmetry of the curves and the periodically going to zero indicate that the gate is properly calibrated. \textbf{b,} Here we demonstrate that we have corrected for changes in the single qubit phase that  result from the interaction gate.  We rotate one qubit to the equator of the Bloch sphere, turn on the interaction for a variable length of time, and then perform state tomography on the qubit which we rotated.  The agreement with theory indicates that the phase is being properly corrected for. \textbf{c,} The relative control phases of the different microwave signals also needs to be corrected for.  Here we rotate one qubit to the equator of the Bloch sphere, turn on the interaction for a variable length of time, and then perform state tomography on a neighboring qubit. The agreement of the curves with theory indicates that we have properly calibrated for this phase difference.
}
\label{fig:GateCalibration} 
\end{figure}

After evolving the system forward $N$ times, we reconstruct the density matrix of the qubits using state tomography.
State tomography consists of single qubit rotations, shown in black, followed by measurements along the z-axis; this is then repeated for various rotation axes and angles.
The rotations are chosen from a set of four rotations containing $I$, $X_{\pi/2}$,  $Y_{\pi/2}$, and  $X_{\pi}$.
The measured z-projections are then used along with maximum likelihood estimates to construct a physical density matrix \cite{neeley2010generation}.

\section{III. Simultaneous three-qubit interaction}

The characterization procedure for the simultaneous three-qubit interaction is shown in Fig.\,\ref{fig:GateCalibration}.
This procedure is broken up into three steps.
First, we calibrate the six square pulse amplitudes (three qubits, three couplers) to ensure that the interaction strengths are all equal and that the qubits are on resonance.
Second, these pulses can cause the qubits to detune from the microwave source; measuring this detuning allows us to correct for the resulting phase accumulation.
Third, if there is a relative phase between the control signals on different qubits, this also needs to be corrected for.

The first experiment, shown in panel (a), demonstrates that the interaction energies are symmetric and that the qubits are on resonance.
We begin by putting one of the qubits into its excited state, turning on the interaction for a variable length of time, and then measuring all three qubit excited state probabilities $P_e$.
This experiment is then repeated exciting a different qubit each time; all 9 curves are plotted as a function of interaction length.
In order to isolate the effects of interaction, we measure $P_e$ as a function of time, without interactions, and subtract the results.
If the qubits are detuned or the interaction strengths differ from one another, then the curves will not lie on top of each other.
Additionally, both error sources will prevent the probabilities from returning to zero periodically.
The data suggests that errors in the coupling and detuning are small over relevant time scales.

\begin{figure}[t!]
\includegraphics{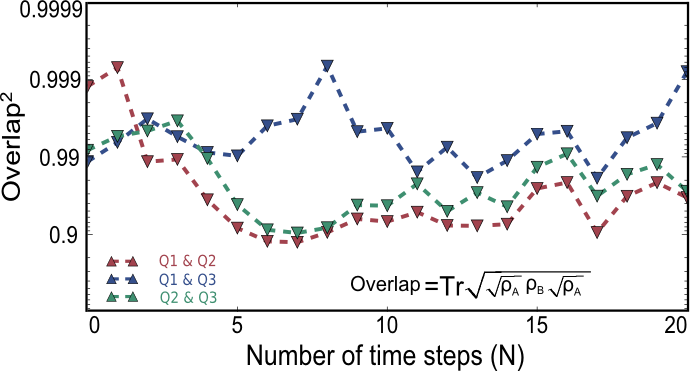} 
\caption {\textbf{Symmetric evolution.} We measure the single qubit density matrices as a function of the number of time steps $N$ for $\kappa = 2.5$.  At each time, we compute the overlap of the individual single qubit density matrices and plot the results.
}
\label{fig:SymmetricEvolution} 
\end{figure}

Measurements of $P_e$ alone do not provide information about the phase of the qubit.
In panel (b), we rotate one qubit to the equator of the Bloch sphere, turn on the interaction for a variable length of time, and then perform state tomography on the qubit which was rotated; we plot the expectation values of the single qubit Pauli operators.
If the qubit is accumulating a phase during the interaction as a result of detuning from the microwave source, then $\expect{x}$ and $\expect{y}$ will rotate into one another.
We determine the rate of phase accumulation by measuring $\expect{y}$ for a 210\,ns interaction length as a function of the phase accumulation rate correction and look for a minimum, as $\expect{y}$ is ideally minimum for this choice of interaction length.
Correcting for this results in tomography which agrees well with an ideal operation;  deviations result primarily from measurement visibility.

In panel (c), we rotate one qubit to the equator of the Bloch sphere, turn on the interaction for a variable length of time, and then perform state tomography on a neighboring qubit.
If the relative phase of the microwave control signals on the individual qubits is non-zero, then the measured $\expect{x}$ and $\expect{y}$ values will rotate into one another.
This may result from differences in electrical path lengths in the two control lines.
We determine this phase by measuring $\expect{y}$ for a 105\,ns interaction length as a function of the relative phase and look for a minimum.
Correcting for this static phase difference results in tomography which agrees well with an ideal operation.

\section{IV. Qubit dynamics}

\begin{figure}[t!]
\includegraphics{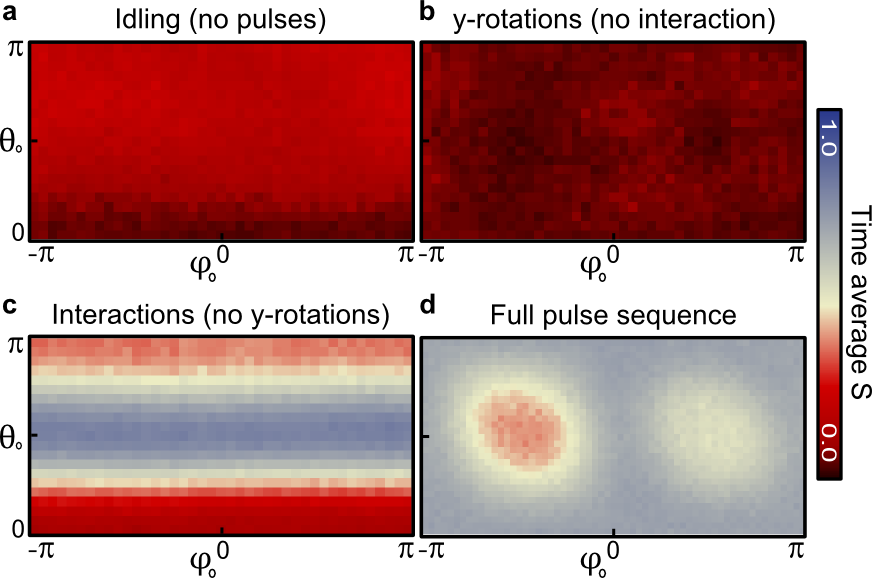} 
\caption {
\textbf{Dissecting the phase space dynamics}. In all four panels we plot the time-average entanglement entropy of a single qubit versus initial state for $N = 20$ and $\kappa = 0.5$.  To better understand the results, we consider four different pulse sequences:  no pulses (just waiting), only $y$-rotations (no interactions), only interactions (no rotations), and both interactions and rotations. \textbf{a,}  Average entanglement entropy after waiting a length of time equivalent to the full pulse sequence.  \textbf{b,} Here, we apply only the $y$-rotations and replace the interactions with a wait of equivalent length. \textbf{c,}  Now we perform the opposite experiment, applying only the interactions and wait instead of rotating.  \textbf{d,}  We now apply the full pulse sequence.
}
\label{fig:PhaseSpaceBrokenUp} 
\end{figure}

\begin{figure}[t!]
\includegraphics{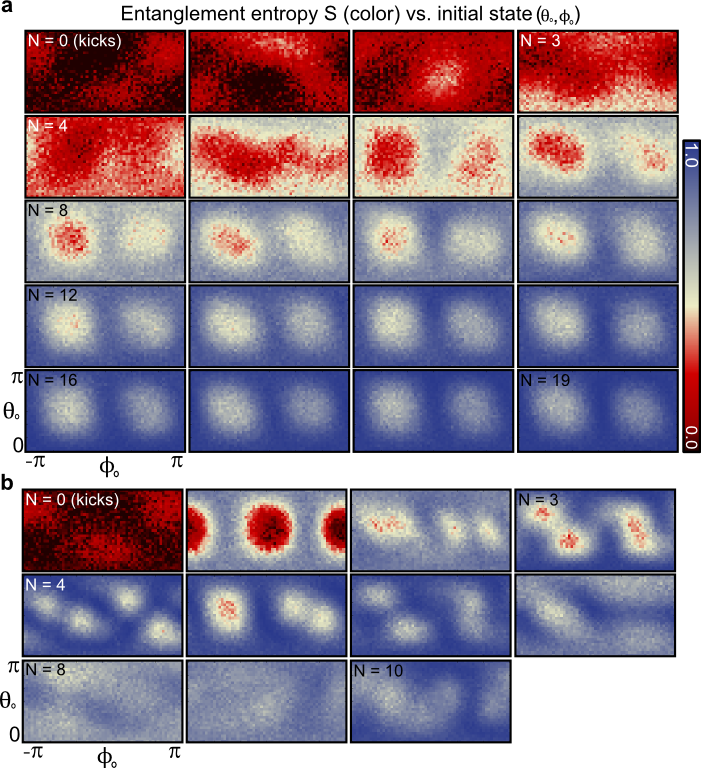} 
\caption {
\textbf{Snapshots of entanglement entropy. a,} Entanglement entropy of a single qubit as a function of intial state for $N = 0$ to $19$ at $\kappa = 0.5$.  \textbf{b,}  We repeat the experiment for $N = 0$ to $10$ at $\kappa = 2.5$.
}
\label{fig:PhaseSpaceEveryStep} 
\end{figure}

This three-qubit interaction along with local rotations are used to generate the dynamics that were explored in this experiment.
As both the initial state and the evolution operators are symmetric under exchange of qubits, we expect to observe nominally identical behavior.
In order to verify this, we measure the reduced density matrix of the individual qubits and compute their overlap.
The results are shown in Fig.\,\ref{fig:SymmetricEvolution} for $\kappa = 2.5$ and an initial state along the $z$-axis.
We find that the qubits remain symmetric over the length of the evolution.

\begin{figure}[t!]
\includegraphics{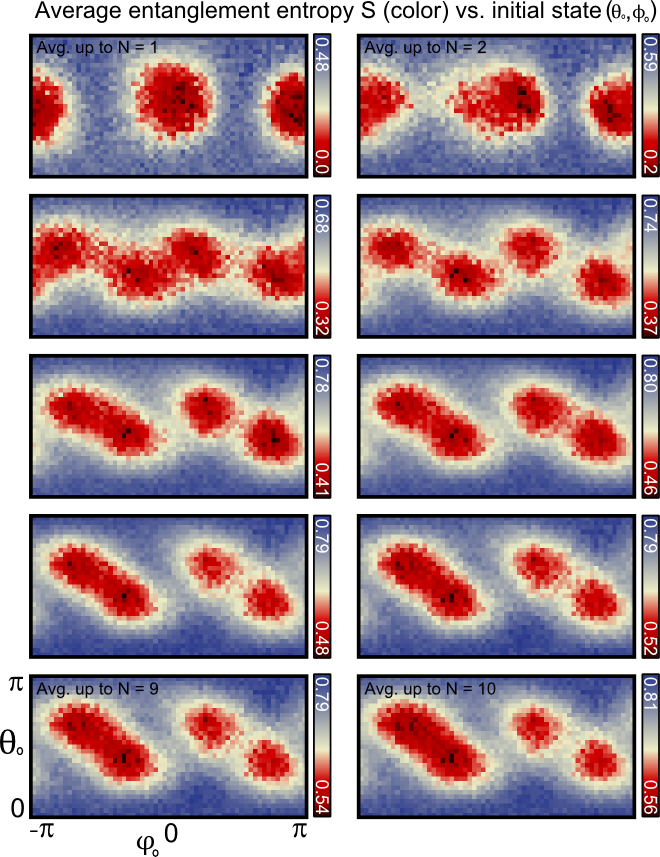} 
\caption {
\textbf{Entanglement entropy, convergence with number of averages.} Entanglement entropy of a single qubit as a function of intial state at $\kappa = 2.5$.  In the different panels, we increase the number of time steps over which we average the entropy.  We find that the entanglement entropy qualitatively converges to the long time behavior after merely four time steps.
}
\label{fig:PhaseSpaceVsAverages} 
\end{figure}

\begin{figure}[t!]
\includegraphics{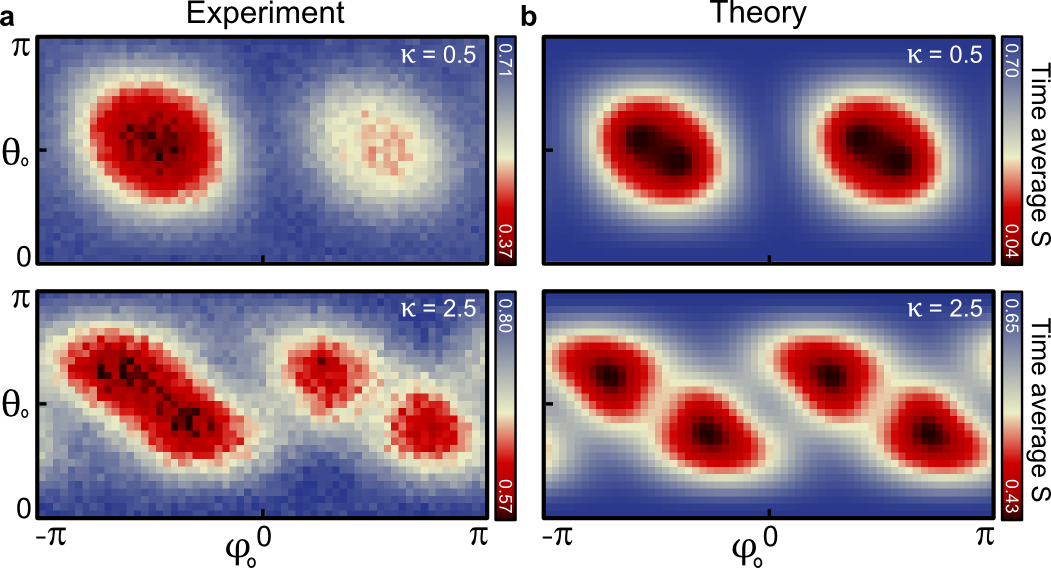} 
\caption {
\textbf{Entanglement entropy, comparison with theory. a,} The time-average entanglement entropy of a single qubit versus intial state for $\kappa = 0.5$ (top) and $\kappa = 2.5$ (bottom) \textbf{b,}  For comparison, we numerically compute the expected behavior and plot the results.
}
\label{fig:PhaseSpaceTheory} 
\end{figure}

\begin{figure}[t!]
\includegraphics{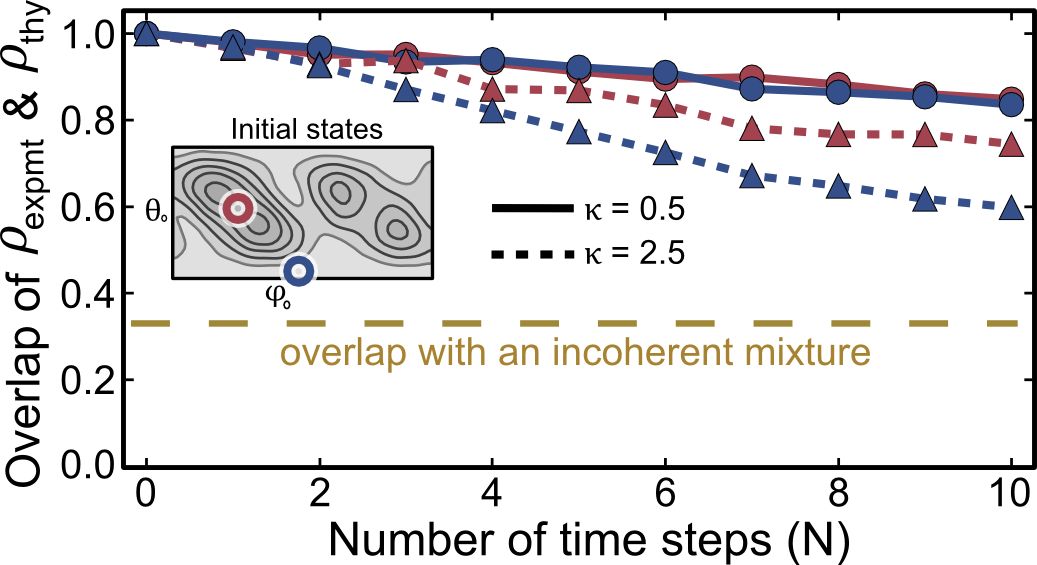} 
\caption {
\textbf{Comparison with theory.}  We measure the three-qubit density matrix for two initial states, one where subsystems thermalized (blue) and one where subsystems did not thermalize (red).  We plot the overlap of the experimental density matrix $\rho_{\eqtext{expmt}}$ and the theoretical density matrix $\rho_{\eqtext{thy}}$ calculated using the model presented in the main text.
}
\label{fig:overlapwiththeory} 
\end{figure}

The evolution of the qubits involves both a rotation and an interaction.
In Fig.\,\ref{fig:PhaseSpaceBrokenUp} we explore the effect of these pulses on the entanglement entropy of the individual qubits.
In panel (a) we plot the time-average entropy versus initial state without either the rotation or interaction; instead, we simply wait for the corresponding length of time.
For initial states near the ground state, the entropy is close to zero and increases slightly while approaching the excited state as a result of energy relaxation.
In panel (b) we plot the same quantity, however, now we apply only the rotations without the interactions.
Here, the entropy is uniform over initial states as the rotations average the results over many states.
In panel (c), we apply only the interaction without the rotations.
We see that near the ground or excited states the entropy stays near-minimum as the qubits do not entangle here.
For initial states closer to the equator, we see an entanglement entropy near a half.
Putting the interaction and the rotation together we recover the results shown in the main text.

In Fig.\,2 of the main text, we show the entanglement entropy at single instances in time for N = 1,3,5 and 7 for both $\kappa = 0.5$ and $\kappa = 2.5$.
In Fig.\,\ref{fig:PhaseSpaceEveryStep}, we show the data for all time steps from $N = 1$ to $20$ for $\kappa = 0.5$ (a) and from $N = 1$ to $10$ for $\kappa = 2.5$ (b).
In Fig.\,\ref{fig:PhaseSpaceVsAverages}, we vary the number of time steps over which we average the entanglement entropy. We find that the regions of high and low entropy qualitatively approach the long time results after just four steps.
In Fig.\,\ref{fig:PhaseSpaceTheory}a, we show the entanglement entropy average over $N$, as shown in the main text.
For comparison, we numerically compute the ideal behavior and show the results in  Fig.\,\ref{fig:PhaseSpaceTheory}b.
The ideal behavior has a left/right symmetry that is not present in the experimental data.
This is likely the result of control errors arising from imperfect calibrations and modifications to the dynamics resulting from dispersive shifts from higher states of the transmon qubit.

\begin{figure}[t!]
\includegraphics{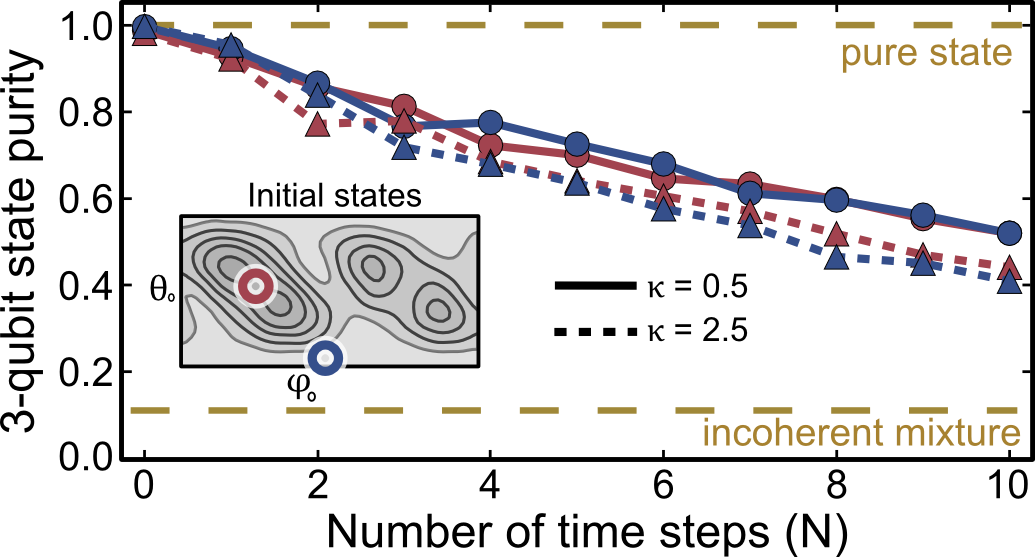} 
\caption {
\textbf{State purity as a measure of decoherence.}  We measure the three-qubit density matrix for two initial states, one where subsystems thermalized (blue) and one where subsystems did not thermalize (red).  We plot the state purity, a measure of decoherence, as a function of the number of time-steps.  We find that the decoherence is independent of initial state for all times and for both values of interaction strength $\kappa = 0.5$ and $\kappa = 2.5$.  This suggests that the contrast between high and low entropy, entanglement and ergodicity found in the main text is the result of coherent quantum dynamics.
}
\label{fig:statepurity} 
\end{figure}

In Fig.\,\ref{fig:overlapwiththeory} we consider the degree to which the model outlined in the main text describes the experimental results.
Using the measured three-qubit density matrix $\rho_{\eqtext{expmt}}$, we compute the overlap of $\rho_{\eqtext{expmt}}$ and the theoretically calculated density matrix $\rho_{\eqtext{thy}}$.
We plot the results as a function of time for two initial states, one where subsystems thermalized (blue) and one where subsystems did not thermalize (red), and for two values of interaction strength, $\kappa = 0.5$ and $\kappa = 2.5$.

\section{V. Unitary dynamics vs. decoherence}
In the main text we show that single-qubit subsystems approach maximal entropy (i.e thermalize, Fig.\,2) as a result of entanglement (Fig.\,3).
Additionally, we show that this occurs for initial states where time-averages are equal to state-space averages (i.e the dynamics are ergodic, Fig.\,4).
In contrast, we find that where the dynamics are less ergodic that subsystems do not thermalize or entangle.
However, we have yet to determine if the contrast between high and low entropy, entanglement, and ergodicity results from unitary dynamics or environmental decoherence.

In Fig.\,\ref{fig:statepurity} we show the state purity of the three-qubit density matrix $\rho_{\eqtext{expmt}}$ as a function of time.
We plot the purity for both an initial state where subsystems thermalized (blue) and did not thermalize (red).
The state-purity, a measure of decoherence, is given by $\Tr{\rho_{\eqtext{expmt}}^2}$ and is $1$ for a pure state and $1/2^3$ for a three-qubit incoherent mixture.
We find that the decoherence is independent of the initial state of the qubits.
This result strongly suggests that the contrast in entropy, entanglement, and ergodicity is the result of coherent quantum dynamics.

\section{VI. Finite-size scaling}

\begin{figure}[t!]
\includegraphics{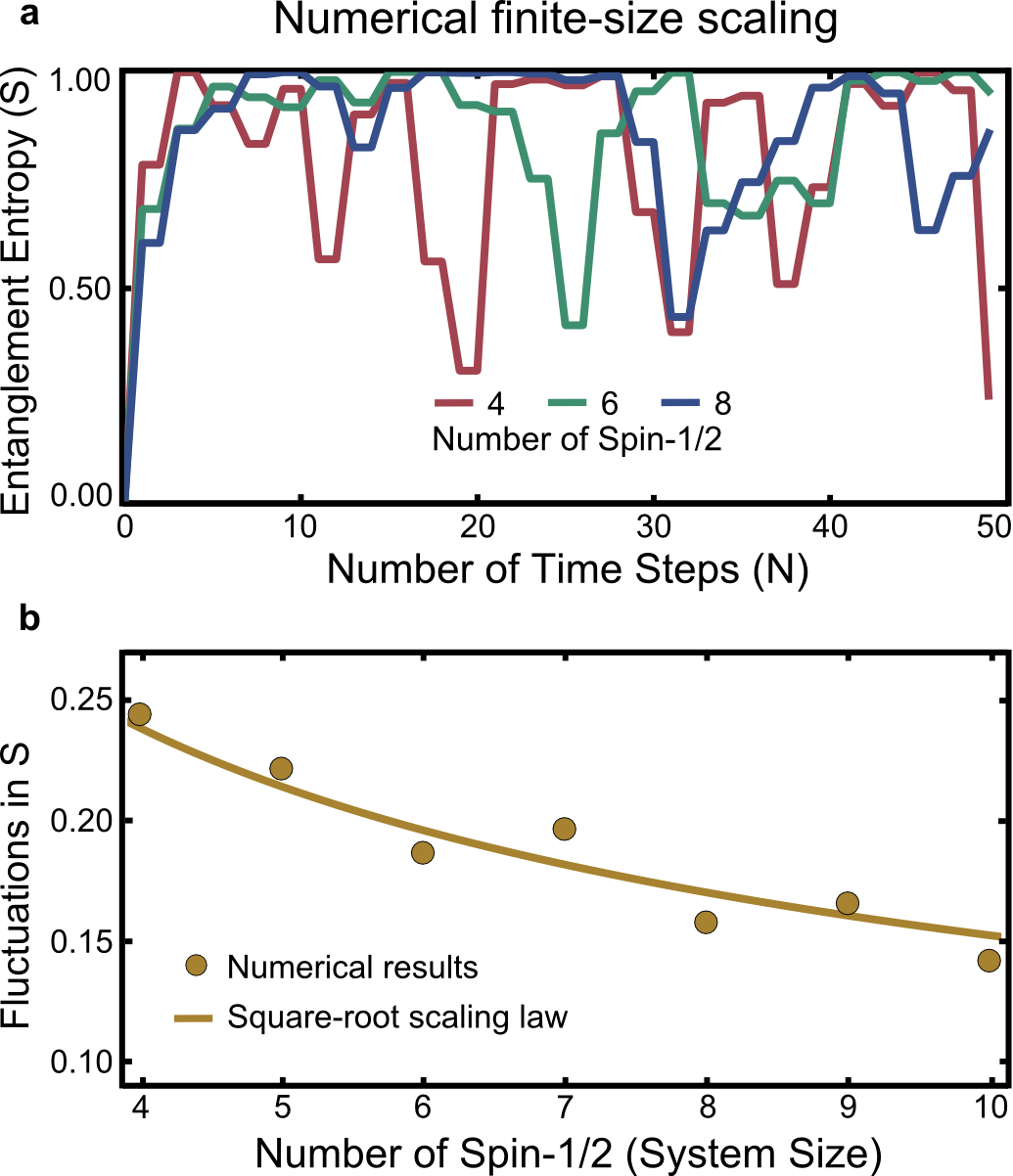} 
\caption {
\textbf{Decreasing fluctuations with system size. a,}  We numerically compute the entanglement entropy S versus the number of time steps N for increasing number of qubits.    In all cases, the entropy approaches 1.0 after a few steps.  However, there are significant fluctuations from this value over time due to the small size of the system. \textbf{b,}  In the lower panel, we numerically compute the standard deviation in entropy from N = 10 to 500 as a function of the number of qubits and show that fluctuations in entropy decrease with increasing system size.  For comparison, we overlay a curve with the square-root of system-size behavior typically found in statistical mechanics.
}
\label{fig:FiniteSizeScaling} 
\end{figure}

In statistical mechanics, fluctuations from equilibrium are expected to vanish with increasing system size.  In our experiment, we average over these fluctuations in order to estimate the equilibrium value of entropy.  In  Fig.\,\ref{fig:FiniteSizeScaling}, we numerically show that these fluctuations in entropy over time decrease as we consider larger systems.  The points correspond to the standard deviation in entropy from N = 10 to 500 as a function of the number of spin-1/2 from 4 to 10.  The solid line corresponds to the expected behavior from statistical mechanics where fluctuations decrease with the square root of system size.  We find agreement between the fluctuations as computed from the quantum dynamics and the predictions from statistical mechanics.

\begin{figure}[t!]
\includegraphics{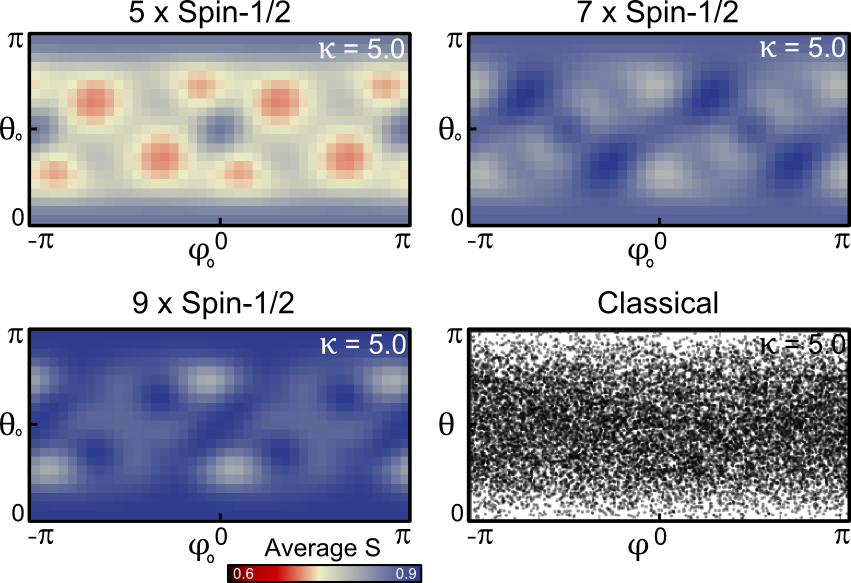} 
\caption {
\textbf{Thermalization for all initial states. }  We numerically compute the time-average entanglement entropy S as a function of initial state for an interaction energy $\kappa = 5.0$.   The value of $\kappa$ is chosen so that the classical phase space is no longer mixed but completely chaotic.  In the first three panels we observe that the time-average entropy increases as a function of the number of spins, for all initial states.  This suggests that at strong interaction all initial states thermalize in the limit of large systems.  In the last panel (lower right), we show the classical phase space dynamics for comparison.
}
\label{fig:AllInitialStatesAppearToThermalize} 
\end{figure}

A major achievement of statistical mechanics is the ability to predict the behavior of physical systems independent of their initial configuration.  In our experiment, we show a clear difference in the entropy of initial quantum states whose classical limits are either chaotic or stable.  If the system were thermal for all initial states, then we would not expect this state-dependent behavior.  In  Fig.\,\ref{fig:AllInitialStatesAppearToThermalize}, we consider larger values of interaction strength where the classical phase space is completely chaotic and compute the quantum evolution. 

When the classical phase space is completely chaotic, we find the the entropy increases with system size independent of the initial state.  This further supports the conclusion  in the matin text that the observations correspond to a thermalization process.

\bibliography{cites}